\documentclass{aa}

\usepackage[varg]{txfonts}
\usepackage{siunitx}
\usepackage[version=4]{mhchem}
\usepackage{multirow}
\usepackage{graphicx}
\usepackage[caption=false]{subfig}
\usepackage[strict]{changepage}
\usepackage{xcolor}
\usepackage{ulem}
\usepackage[colorlinks=true,citecolor=blue]{hyperref}
\usepackage{float}
\usepackage{amssymb}
\usepackage{amsmath}
\usepackage{lscape}
\usepackage{mathtools}
\usepackage{natbib}
\usepackage{pdflscape}
\usepackage{gensymb}

\defcitealias{zhou2023}{Paper~I}   

\DeclareSIUnit\jansky{Jy}

\newcommand{\msun}{$\rm M_\odot$}

\newcolumntype{L}[1]{>{\raggedright\let\newline\\\arraybackslash\hspace{0pt}}m{#1}}
\newcolumntype{C}[1]{>{\centering\let\newline\\\arraybackslash\hspace{0pt}}m{#1}}
\newcolumntype{R}[1]{>{\raggedleft\let\newline\\\arraybackslash\hspace{0pt}}m{#1}}

\begin{document}

\title{High-resolution APEX/LAsMA $^{12}$CO and $^{13}$CO (3-2) observation of the G333 giant molecular cloud complex : II.  Survival and gravitational collapse of dense gas structures under feedback}
\author{J. W. Zhou\inst{\ref{inst1}} 
\and F. Wyrowski\inst{\ref{inst1}} 
\and S. Neupane\inst{\ref{inst1}}
\and I. Barlach Christensen\inst{\ref{inst1}}
\and K. M. Menten
\inst{\ref{inst1}}
\and S. H. Li\inst{\ref{inst2}}
\and T. Liu\inst{\ref{inst3}} 
}
\institute{Max-Planck-Institut f\"{u}r Radioastronomie, Auf dem H\"{u}gel 69, 53121 Bonn, Germany \label{inst1} \\
\email{jwzhou@mpifr-bonn.mpg.de}
\and
Max Planck Institute for Astronomy, K\"{o}nigstuhl 17, 69117 Heidelberg, Germany \label{inst2}
\and
Shanghai Astronomical Observatory, Chinese Academy of Sciences, 80 Nandan Road, Shanghai 200030, Peoples Republic of China \label{inst3}
}

\date{Accepted XXX. Received YYY; in original form ZZZ}

\abstract
{Feedback from young massive stars has an important impact on the star formation potential of their parental molecular clouds.}
{We investigate the physical properties of gas structures under feedback in the G333 complex using data of the $^{13}$CO $J= 3-2$ line observed with the LAsMA heterodyne camera on the APEX telescope.}
{We used the Dendrogram algorithm to identify molecular gas structures based on the integrated intensity map of the $^{13}$CO (3$-$2) emission, and extracted the average spectra of all structures to investigate their velocity components and gas kinematics.}
{We derive the column density ratios between different transitions of the $^{13}$CO emission pixel-by-pixel, and find the peak values $N_{\rm 2-1}/N_{\rm 1-0} \approx$ 0.5, $N_{\rm 3-2}/N_{\rm 1-0} \approx$ 0.3, $N_{\rm 3-2}/N_{\rm 2-1} \approx$ 0.5. These ratios can also be roughly predicted by the non-LTE molecular radiative transfer code RADEX for an average H$_2$ volume density of $\sim 4.2 \times 10^3$ cm$^{-3}$. 
A classical virial analysis does not reflect the true physical state of the identified structures,
and we find that external pressure from the ambient cloud plays an important role in confining the observed gas structures. 
For high column density structures, velocity dispersion and density show a clear correlation, while for low column density structures they do not, indicating the contribution of gravitational collapse to the velocity dispersion. Branch structures show a more significant correlation between 8 µm surface brightness and velocity dispersion than leaf structures, implying that feedback has a greater impact on large-scale structures. 
For both leaf and branch structures, $\sigma-N*R$ always has a stronger correlation compared to $\sigma-N$ and $\sigma-R$. 
The scaling relations are stronger, and have steeper slopes when considering only self-gravitating structures, which are the structures most closely associated with the Heyer-relation.}
{Although the feedback disrupting the molecular clouds will break up the original cloud complex, the substructures of the original complex can be reorganized into new gravitationally governed configurations around new gravitational centers. This process is accompanied by structural destruction and generation, and changes in gravitational centers, but gravitational collapse is always ongoing.}

\keywords{Submillimeter: ISM -- ISM:structure -- ISM: evolution -- Stars: formation -- methods: analytical -- techniques: image processing}

\titlerunning{Gravitational collapse of dense gas structures under feedback}
\authorrunning{J. W. Zhou}

\maketitle 

\section{Introduction}
High-mass stars (M$\textgreater$8 M$_{\odot}$) have a profound impact on the evolution of the interstellar medium (ISM). Throughout their short lifetimes ($\sim$10$^{6}$ yr),  radiation-driven stellar winds from high-mass stars create HII regions in the surrounding giant molecular clouds (GMCs) \citep{Zinnecker2007-45, Molinari2014, Motte2018-56}. 
High-mass stars end their lives in the form of supernovae (SNe) whose explosions can 
release $\sim$10$^{51}$ erg of energy near-instantaneously. Shocks from expanding HII regions and supernova remnants (SNRs) can accelerate and heat their surrounding gas, and add turbulence to the gas. In simulations, massive stellar feedback, including ionizing radiation, stellar winds and supernovae
\citep{Matzner2002-566, Dale2012-424, Rogers2013-431, Dale2014-442, Rahner2017-470, Smith2018-478, Lewis2023-944}, can suppress the star formation and destroy the natal cloud. However,
whether the stellar feedback promotes or suppresses star formation remains controversial.

W49A is one of the most massive and luminous young star-forming regions in the Galaxy. 
As presented in \citet{Rugel2019-622}, it is more likely that only limited parts of W49A were affected by feedback from the central stellar cluster, while stars in the outer parts of W49A formed independently. Moreover, all feedback models used in \citet{Rugel2019-622} predict re-collapse of the shell after the first star formation event, which means that feedback of the first formed cluster is therefore not strong enough to disperse the cloud.
Previous work on the G305 region observed with the Large APEX sub-Millimeter Array (LAsMA) 7 beam receiver on the Atacama Pathfinder Experiment 12 meter submillimeter telescope (APEX) found that strong stellar winds drive turbulence in the G305 GMC in which  feedback has triggered star formation by the collect and collapse mechanism \citep{Mazumdar2021-650,Mazumdar2021-656}. 
The dense molecular gas structures inside the cloud serve as star-forming sites and their physical states  directly determine the star formation capability of the molecular cloud under feedback. A basic question is how the dense gas structures survive and maintain star formation activity in a strong feedback environment, which depends on the relative strength between their gravity and turbulence.

The relative importance of turbulence and gravity in massive star-forming regions is a long and widely debated topic, distinct views lead to different physical pictures of massive star formation, such as turbulent-core model \citep{McKee2003,Krumholz2007},
competitive-accretion model \citep{Bonnell1997,Bonnell2001}, inertial-inflow model \citep{Padoan2020}, and global hierarchical collapse model \citep{Vazquez2009,Ballesteros2011,Hartmann2012,Vazquez2017,Vazquez2019}.
Larson's laws claim that in molecular clouds the velocity dispersion $\sigma$ scales proportionally to the scale $R$, and molecular clouds are approximately in virial equilibrium, with a mostly uniform column density. The Larson-relation ($\sigma_v \propto R^{0.5}$) is generally used to emphasize the importance of turbulence in molecular clouds, and turbulence acts to sustain the clouds against gravitational collapse \citep{Larson1981-194,Solomon1987-319,Heyer2004-615,MacLow2004-76, McKee2007-45,Hennebelle2012-20}. 
\citet{Heyer2009-699} generalized the Larson-relation by extending the Larson-ratio $L \equiv \sigma_v/R^{0.5}$ with the surface density $\Sigma$ of Galactic GMCs, i.e. $\sigma_v/R^{0.5} \propto \Sigma^{0.5}$. 
Subsequently, the Heyer-relation has been used to emphasize the importance of gravity in molecular clouds \citep{Ballesteros2011-411, Traficante2018-473,Traficante2018-477, Ballesteros2018-479, Vazquez2019-490, Ballesteros2020-216}.
Especially, in the high-column density portions of star-forming regions, such as clumps or cores, the Heyer-relation always performs better than the Larson-relation 
\citep{Ballesteros2011-411,Camacho2016-833, Traficante2018-477}, suggesting strong gravity at relatively small scales in molecular clouds. \citet{Ibanez2016-824} have shown that the Heyer-relation cannot be reproduced without self-gravity in simulations of the ISM with supernova-driven turbulence. In contrast, purely supernova-driven turbulence in the ISM generates the Larson-relation.

Regarding the explanation of the Heyer-relation, \citet{Heyer2009-699} claimed that it is consistent with the clouds being in virial equilibrium, as it follows directly from the condition $2E_{\rm k} = E_{\rm g}$, where $E_{\rm k} = M \sigma^2/2$, $E_{\rm g} = 3 GM^2/5R$. \citet{Ballesteros2011-411} further pointed out that the scaling is also consistent with the clouds undergoing free-fall, in which case $E_{\rm k} = |E_{\rm g}|$. However, the differences between the effects of free-fall and virial equilibrium in the $\sigma_v/R^{0.5}$ vs.\ $\Sigma$ diagram are smaller than the typical uncertainty of the observational data \citep{Ballesteros2011-411}, thus difficult to distinguish.
Both explanations involve only gravitational and kinetic energy, which may be a workable approximation for relatively isolated molecular clouds, but is often too simplistic for substructures inside a molecular cloud, especially for a cloud affected by feedback, such as our target G333. When the substructures can self-gravitationally collapse, they may decouple from the surrounding environment \citep{Peretto2023-525}. If not, the exchange of energy with the surrounding environment will break the conversion between gravitational potential energy, E$_{g}$, and kinetic energy, E$_{k}$, of the structures, and thus violate the Heyer-relation.

In \citet{Zhou2023-676}, we found in the G333 complex that the larger scale inflow is driven by the larger scale cloud structure, indicating hierarchical structure in the GMC and gas inflow from large to small scales. The large-scale gas inflow is driven by gravity, implying that the molecular clouds in the G333 complex may be in a state of global gravitational collapse. However, the broken morphology of some very infrared bright structures in the G333 complex also indicates that feedback is disrupting star-forming regions. 
Here we are going to address the question of how the dense molecular structures survive and maintain the gravitational collapse state in a strong feedback environment.

\section{Data}
\subsection{LAsMA data}
The observations and
data reduction have been described in detail in \citet{Zhou2023-676}.
We mapped a $3.4\degr \times 1.2\degr$ area centered at $(l,b)=(332.33\degr,-0.29\degr)$ using the APEX telescope \citep{Gusten2006-454}.
\footnote{\tiny This publication is based on data acquired with the Atacama Pathfinder EXperiment (APEX). APEX is a collaboration between the Max-Planck-Institut f\"{u}r Radioastronomie, the European Southern Observatory and the Onsala Space Observatory.} 
The 7 pixel Large APEX sub-Millimeter Array (LAsMA) receiver was used to observe the $J=3-2$ transitions of $^{12}$CO ($\nu_{\text{rest}}\sim345.796$ \rm GHz) and $^{13}$CO ($\nu_{\text{rest}}\sim 330.588$ \rm GHz) simultaneously. 
The local oscillator frequency was set at 338.190 \rm GHz in order to avoid contamination of the $^{13}$CO (3$-$2) spectra due to bright $^{12}$CO (3$-$2) emission from the image side band.
Observations were performed in a position switching on-the-fly (OTF) mode. 
The data were calibrated using a three load chopper wheel method, which is an extension of the ''standard'' method used for millimeter observations \citep{Ulich1976-30} to calibrate the data in the corrected antenna temperature $T^*_A$ scale. The data were reduced using the GILDAS package\footnote{\url{http://www.iram.fr/IRAMFR/GILDAS}}. The final data cubes have a velocity resolution of 0.25 km s$^{-1}$, an angular resolution of $19.5 \arcsec$ and a pixel size of $6 \arcsec$.
A beam efficiency value $\eta_{mb}=0.71$ \citep{Mazumdar2021-650} was used to convert intensities from the $T^*_A$ scale to main beam brightness temperatures, $T_{mb}$.

\subsection{Archival data}\label{archive}
To allow for column density estimates using different $^{13}$CO transitions, we also collect  $^{13}$CO (1$-$0) data from the Mopra-CO survey \citep{Burton2013-30} and $^{13}$CO (2$-$1) from the SEDIGISM survey \citep{Schuller2021-500}. 
However, the two surveys only cover galactic latitudes within $\pm 0.5 \degr$, while our LAsMA observations cover the latitudes range of  $\sim -0.29 \pm 0.6 \degr$. Thus we only consider the overlap region of the three surveys.
The data were smoothed to a common  angular resolution $\sim$35$\arcsec$ and a velocity resolution $\sim$0.25 km s$^{-1}$.

The observed region was covered in the infrared range by the Galactic Legacy Infrared Midplane Survey \citep[GLIMPSE,][]{Benjamin2003}.
The GLIMPS images, obtained with the Spitzer Infrared Array Camera (IRAC) at 4.5 and 8.0 $\mu$m, were retrieved from the Spitzer archive. The angular resolution of the images in the IRAC bands is $\sim 2\arcsec$.
We also used  870 $\mu$m continuum data from the APEX Telescope Survey of the Galaxy \citep[ATLASGAL,][]{Schuller2009-504} combined with lower resolution data from the Planck spacecraft, which are sensitive to a wide range of spatial scales at a resolution of $\sim21\arcsec$ \citep{Csengeri2016-585}. Furthermore, we use Hi-GAL
data \citep{Molinari2010} processed using the PPMAP procedure \citep{Marsh2015-454}, which provides column density and dust temperature maps with a resolution of $\sim12\arcsec$  \citep[the maps are available online\footnote{http://www.astro.cardiff.ac.uk/research/ViaLactea/},][]{Marsh2017-471}.

\section{Results}

\subsection{Dendrogram structures}\label{Dendrogram}

\begin{figure}[htbp!]
\centering
\includegraphics[width=0.5\textwidth]{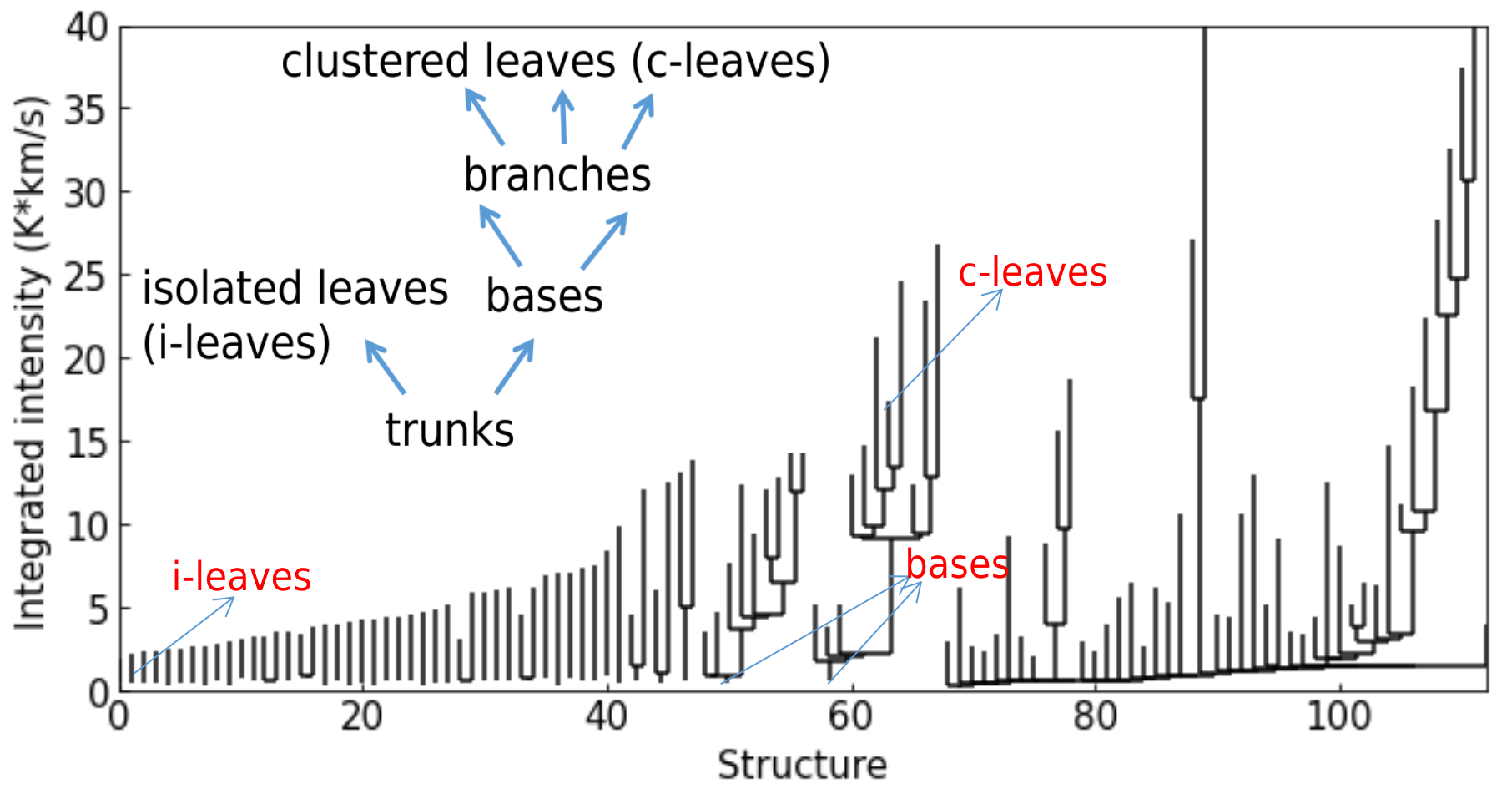}
\caption{Hierarchical structures identified by the Dendrogram algorithm. A segment of the Dendrogram tree of sub-region S3 in Fig.\ref{total}(c) is used to illustrate the  structure types output by the Dendrogram algorithm.}
\label{relation}
\end{figure}

\begin{figure*}[htbp!]
\centering
\includegraphics[width=0.9\textwidth]{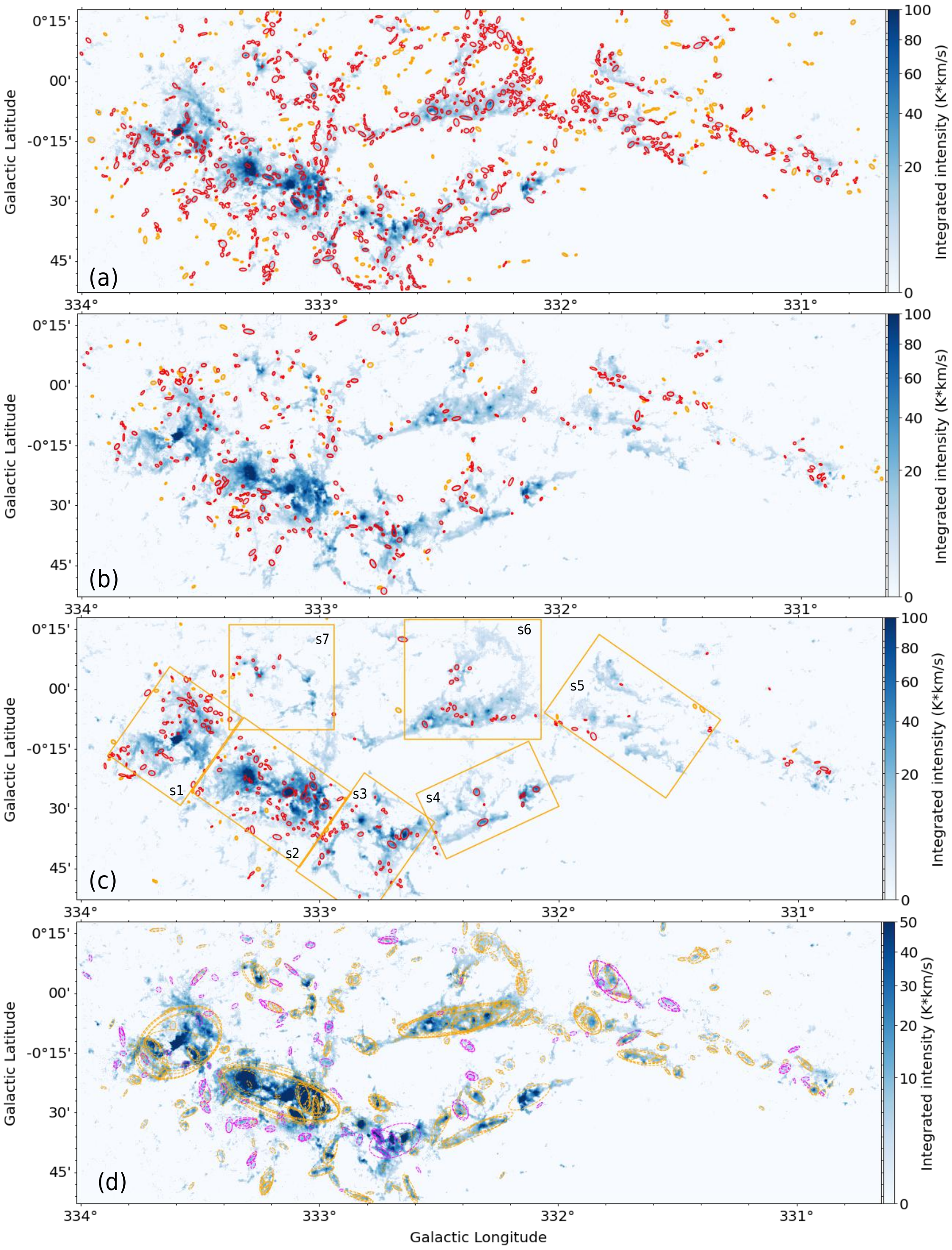}
\caption{Different kinds of structures traced by $^{13}$CO (3$-$2) emission classified in Sec.\ref{components}.
(a) Type1 (single velocity component) leaf structures; (b) Type2 (separated velocity components) leaf structures; (c) Type3 (blended velocity components) leaf structures. Orange boxes mark the sub-regions divided in \citet{Zhou2023-676}; (d) Type1 (orange) and type2 (magenta) branch structures. In panels (a), (b) and (c), orange and red ellipses represent i-leaves and c-leaves, respectively.}
\label{total}
\end{figure*}

\begin{figure}[htbp!]
\centering
\includegraphics[width=0.45\textwidth]{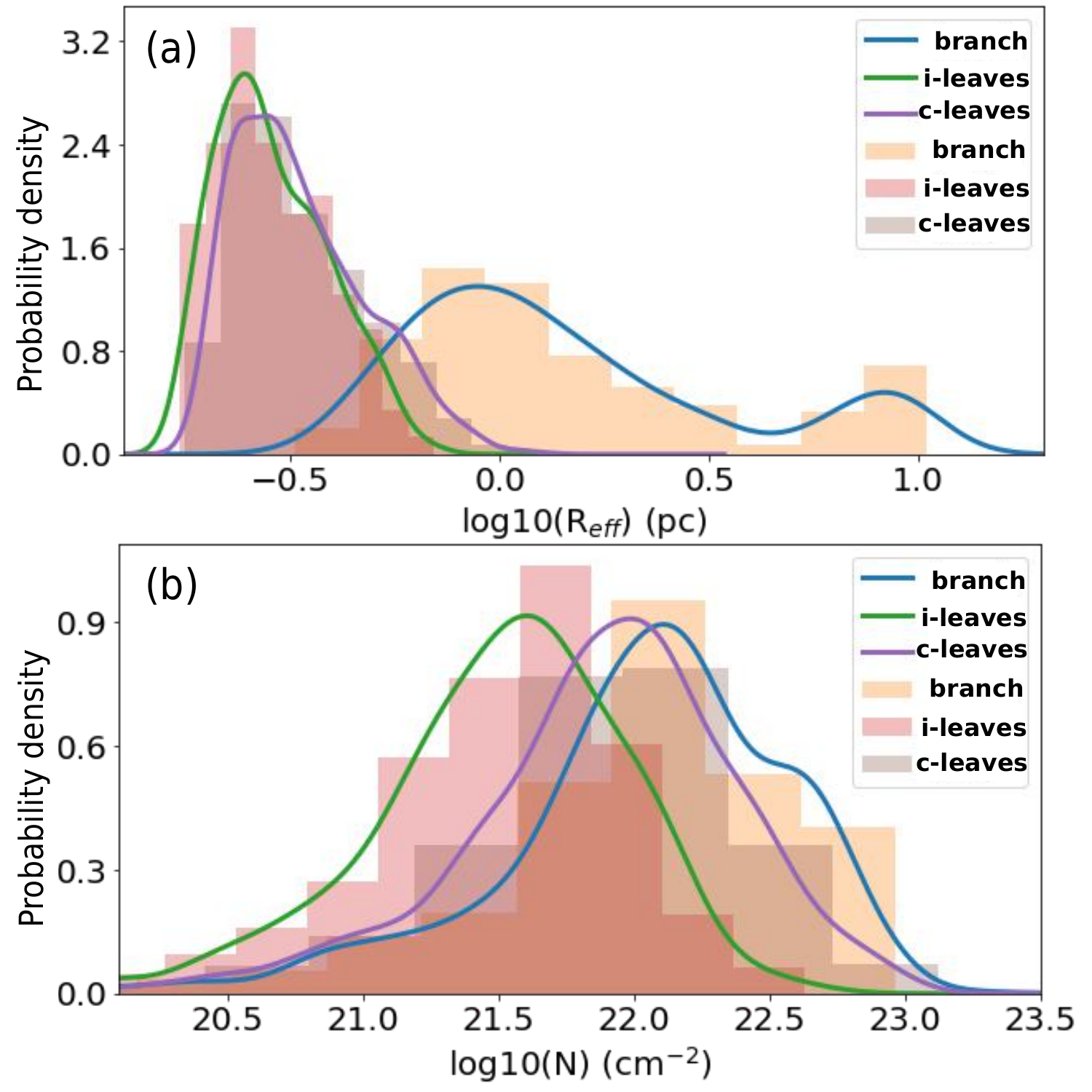}
\caption{(a) Effective radius (scale) and (b) Column density distribution of Dendrogram structures. Only the type1 and type2 structures (see Fig.\ref{total}) are included in the distributions. The probability density is estimated by the kernel density estimation (KDE) method.}
\label{basic}
\end{figure}

\begin{figure*}[htbp!]
\centering
\includegraphics[width=0.85\textwidth]{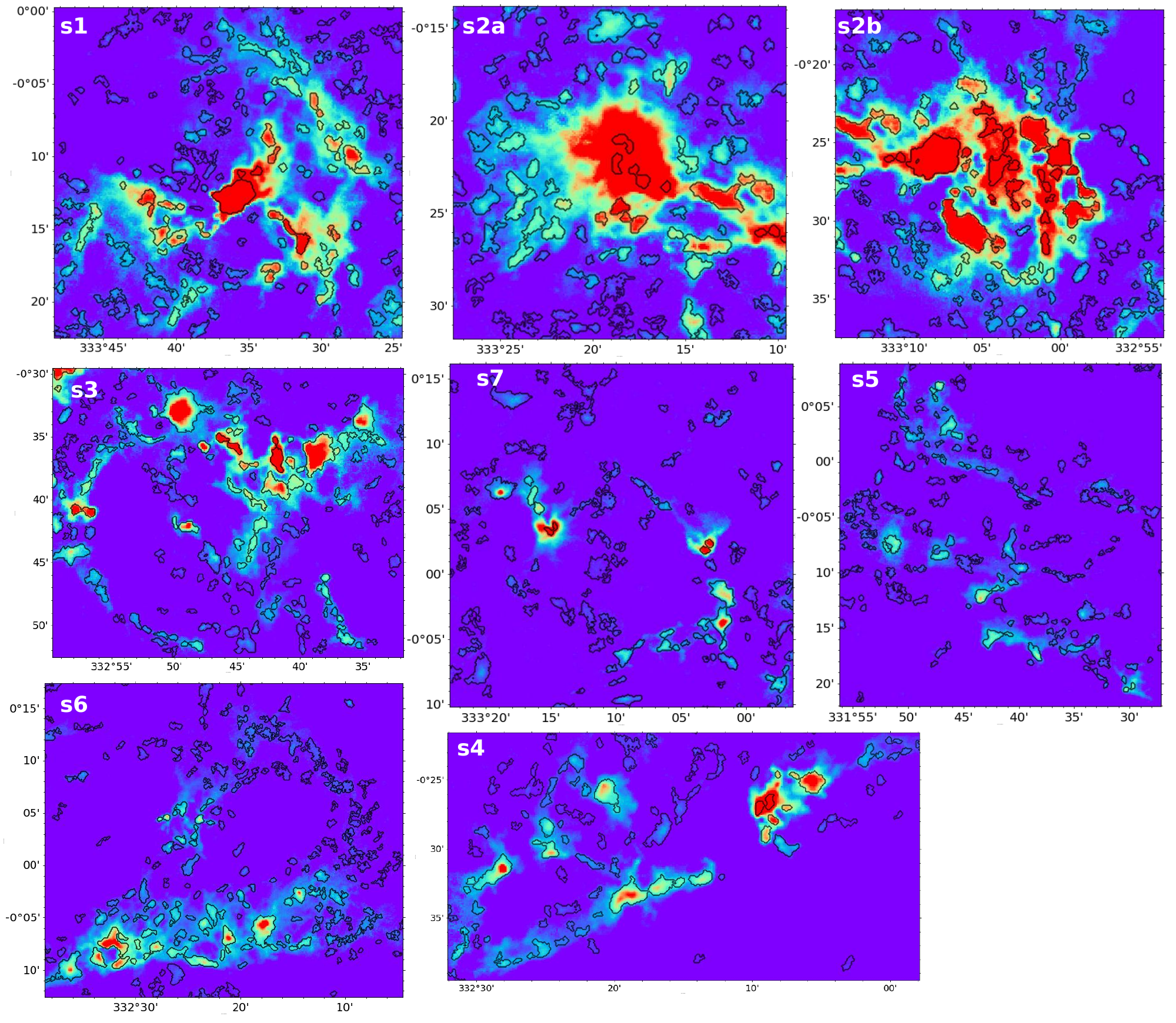}
\caption{Masks of leaf structures identified by the Dendrogram algorithm toward the sub-regions marked in Fig.\ref{total}(c). Only leaf structures are shown in here.}
\label{sub}
\end{figure*}

\begin{figure*}[htbp!]
\centering
\includegraphics[width=0.85\textwidth]{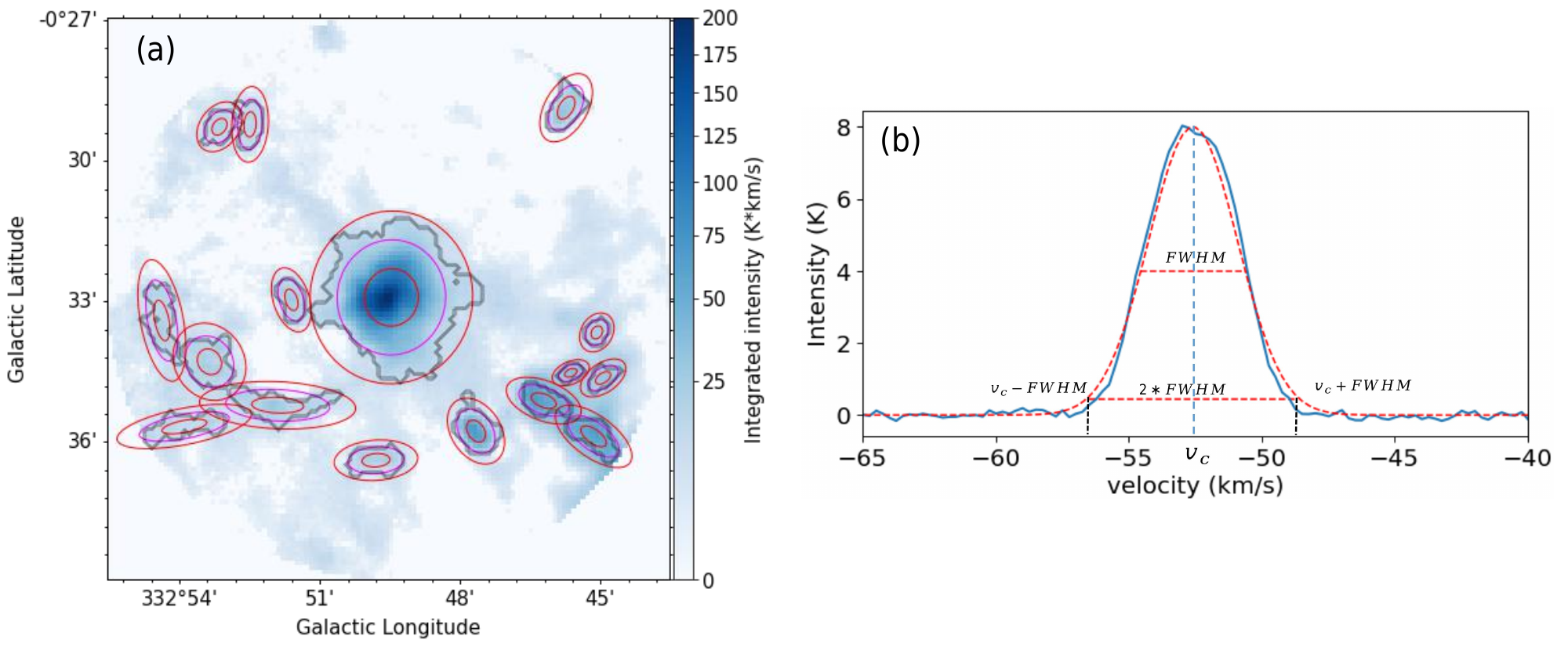}
\caption{A piece of sub-region S3 in Fig.\ref{total}(c) is used to illustrate the  structures identified by the Dendrogram algorithm.
(a) The black contours show the masks of Dendrogram leaves. 
The long and short axes of the smallest ellipse $a$ and $b$ are the rms sizes (second moments) of the intensity distribution. The ellipses in the second and third layers are enlarged by factors of 2 and 3 in size compared to the smallest one. The middle ellipse
visibly corresponds best to the mask; (b) Typical line profile of a leaf structure. The velocity range of the structure is [v$_{c}$-FWHM,v$_{c}$+FWHM].}
\label{example}
\end{figure*}

We identify dense gas structures using the Dendrogram algorithm. As described in \citet{Rosolowsky2008-679}, the Dendrogram algorithm decomposes intensity data (a position-position map or a position-position-velocity cube) into hierarchical structures called leaves, branches and trunks. The relationship between those structures is shown in Fig.\ref{relation}. Trunks are the largest continuous structures at the bottom of hierarchical structures ("bases"), but, by definition, they can also be isolated leaves ("i-leaves") without any parent structure. 
Thus there are two kinds of "trunks", they are called "bases" and "i-leaves" in this work.
Clustered leaves ("c-leaves") are defined as small-scale, bright structures at the top of the tree that do not decompose into further substructures, they are the smallest structures inside "branches". Branches are the relatively large scale structures in the tree, and they can be broken down into substructures. Between  "bases" and "c-leaves", all hierarchical substructures are "branches", thus branches can span a wide range of scales. When we treat "bases" as the largest branches, and combine c-leaves and i-leaves, then there are only two kinds of structures (i.e. leaves and branches). However, in some cases, it is necessary to differentiate between i-leaves and c-leaves. In general, c-leaves are concentrated in regions of relatively high column density, i-leaves are low column density structures distributed at the periphery, as shown in Fig.~\ref{total}. There are no definite limits to the size of the structures at different levels. The size of a leaf structure in a low column density region may be larger than a branch structure in a high column density region. As shown in Fig.\ref{basic}(a), there is considerable overlap of the scales between leaf and branch structures. In general, branches are larger scale structures than leaves. The physical properties of the overlapping parts of leaf and branch structures may be similar, we should remember the mixing in the discussion of the scaling relations below.

Using the {\it astrodendro} package \footnote{\url{https://dendrograms.readthedocs.io/en/stable/index.html}},
there are three major input parameters for the Dendrogram algorithm: {\it min\_value} for the minimum value to be considered in the dataset, {\it min\_delta} for a leaf that can be considered as an independent entity, {\it min\_npix} for the minimum area of a structure.
From these parameters, we can see that the algorithm does not consider the velocity component and velocity range of the identified structure carefully. The structure is mainly identified according to the intensity, thus the velocity division of a structure is only a result of its intensity division. There is no criterion for a continuous velocity range across a dense structure. However, the velocity range of a structure is crucial for the estimation of its fundamental physical quantities, such as velocity dispersion and mass. Moreover, strict differentiation of the velocity components should be based on the spectral line profiles rather than the intensity thresholds in the algorithm.
In this work, instead of identifying structures in the PPV cube, we first identify the intensity peaks on the integrated intensity (Moment 0) map of $^{13}$CO (3$-$2) emission, and then extract the average spectrum of each structure to investigate their velocity components and gas kinematics. 

For the Moment 0 map, a 5$\sigma$ threshold has been set, so we therefore only require the smallest area of a structure to be larger than one beam and do not set any other parameters, thereby reducing the dependence of structure identification on the algorithm parameters. In Fig.~\ref{sub}, the structures identified by the Dendrogram algorithm correspond well to the peaks on the integrated intensity maps.
In order to retain as many structures as possible, the parameter {\it min\_npix} was set to one beam, because the hierarchical structures in Dendrogram mean that a leaf structure under strict parameter settings can be a branch structure under loose parameter settings. Moreover, the average spectra fitting described below will also further screen the structures, allowing to exclude structures with poorly defined line profiles.

The algorithm approximates the morphology of each structure as an ellipse, which is used in this work. We do not use the mask output by the algorithm because different parameter settings around the intensity peak will give different masks. 
In the Dendrogram algorithm, the long and short axes of an ellipse $a$ and $b$ are the rms sizes (second moments) of the intensity distribution along the two spatial dimensions. 
However, as shown in Fig.\ref{example}, $a$ and $b$ will give a smaller ellipse, compared to the size of the identified structure. Thus we tried to enlarge the ellipse by 2 and 3 times, and found that multiplying a factor of 2 is appropriate, similar to a factor of 1.91 suggested in \citet{Solomon1987-319,Rosolowsky2006-118}. 
Then the effective physical radius of an ellipse is $ R\rm_{eff}$ =$\sqrt{2a \times 2b}*d$, here $d=3.6~{\rm kpc}$ for the distance to the G333 complex \citep{Lockman1979-232,Bains2006-367}.

\subsection{Velocity components}\label{components}

\begin{figure}
\centering
\includegraphics[width=0.45\textwidth]{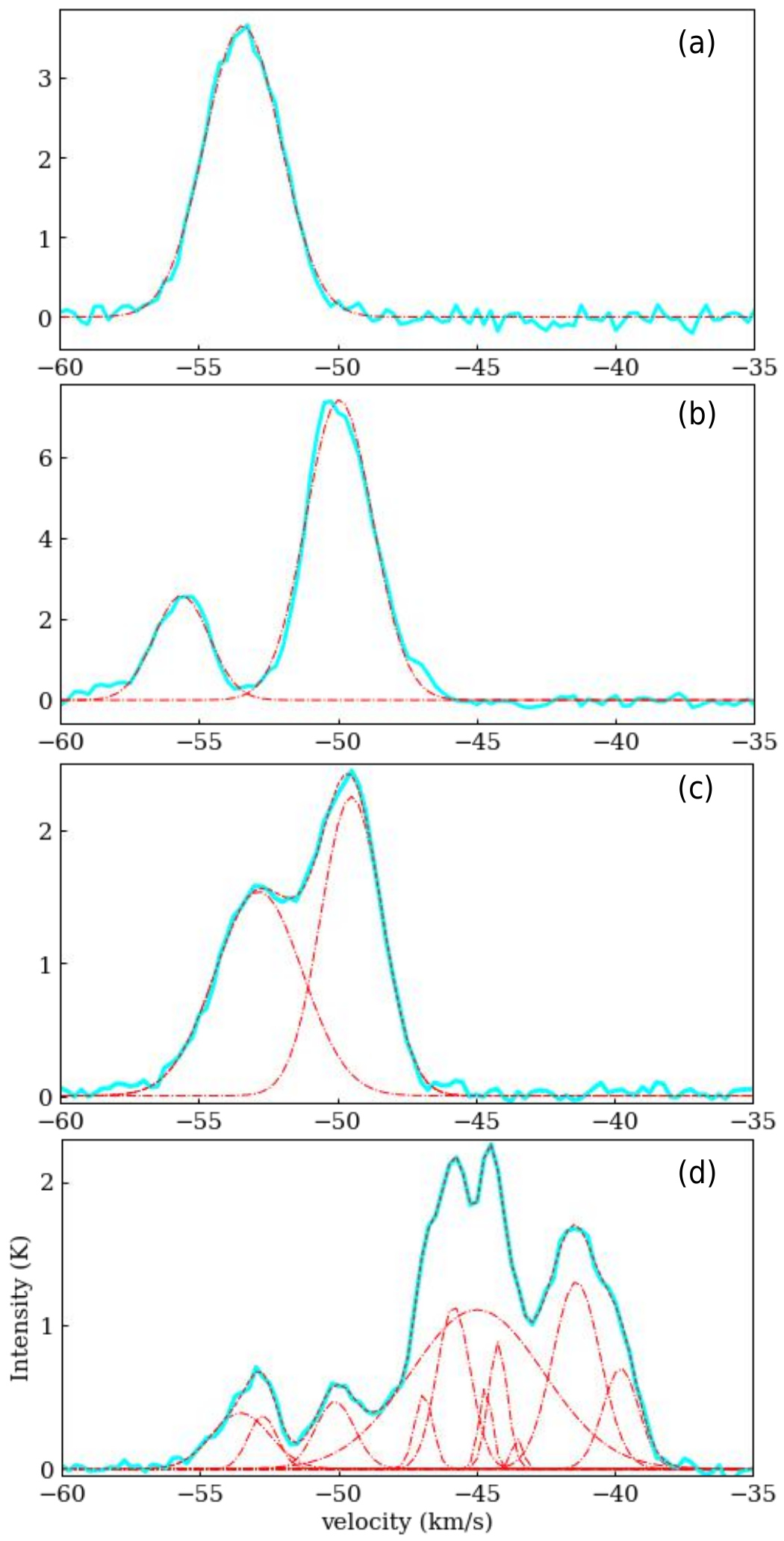}
\caption{Typical $^{13}$CO (3$-$2) line profiles of Dendrogram structures.}
\label{typical}
\end{figure}

Based on the Moment 0 map of the $^{13}$CO (3$-$2) emission, 3608 structures are extracted by the Dendrogram algorithm, consisting of 1626 clustered leaves, 1367 branches and 615 trunks (486 isolated leaves and 129 bases). In the discussion below, we put bases into branches.
We extract and fit the averaged spectra of 3608 structures individually using the fully automated Gaussian decomposer \texttt{GAUSSPY+} \citep{Lindner2015-149, Riener2019-628}. The parameter settings of the decomposition are the same as in \citet{Zhou2023-676}. According to the line profiles, all averaged spectra are divided into three categories:

1. Structures with single velocity components regarded as independent structures (type1, single, 65\%, Fig.\ref{typical}(a));

2. Structures with more than one peak, which are separated (type2, separated, 19\%, Fig.\ref{typical}(b));

3. Structures with more than one peak, blended together (type3, blended, 16\%, , Fig.\ref{typical}(c) and (d)).

Spectra averaged across regions which show a single peak in their line profiles probably represent independent structures. From the line profile, we can also determine the complete velocity range of a structure. In order to ensure that other physical quantities (such as column density, temperature) match with the fitted line-width, as shown in Fig.\ref{example}, we take the velocity range of each structure or each velocity component as [v$_{c}$-FWHM,v$_{c}$+FWHM], which is necessary to calculate the physical quantities for structures with more than one velocity component.

For type3 structures with significant overlapping velocity components, complete decomposition cannot easily be obtained, thus the decomposition uncertainties directly affect the reliability of the subsequent analysis. In this work, we focus on the structures with independent line profiles (type1 and type2). As shown in Fig.\ref{total},
a high-column density structure does not necessarily imply a complex line profile, thus discarding type3 structures won't produce significant sample bias. It is also important to emphasize that for any analysis involving continuum emission without velocity information, only type1 structures will be considered, and sub-regions 5 and 7 marked in Fig.\ref{total}(c) will also be excluded due to the heavy blending of velocity components described in \citet{Zhou2023-676}.

For a type2 structure, we determine the physical size scales of different velocity components based on their velocity ranges.
In Fig.\ref{typical}(b), the total velocity range for deriving the Moment 0 map of the structure is [-60, -35] km s$^{-1}$, the area of a type2 structure on Moment 0 map is $s$ and includes $n$ pixels. For two velocity components in a type2 structure, we can also obtain their Moment 0 maps $m01$ and $m02$ in their velocity ranges [v$_{c1}$-FWHM$_{1}$,v$_{c1}$+FWHM$_{1}$] and [v$_{c2}$-FWHM$_{2}$,v$_{c2}$+FWHM$_{2}$], respectively. $m01$ and $m02$ contain $n1$ and $n2$ pixels, then their area are $(n1/n)*s$ and $(n2/n)*s$, which can be used to estimate the physical size scales of the two velocity components.

Generally, the elliptic approximation for the identified structures is good for small-scale leaf structures, but for some large-scale branch structures, due to their complex morphology, it cannot be satisfactory. We therefore exclude branch structures with complex morphology, if the proportion of empty pixels within the effective ellipse of each structure on the Moment 0 map is larger than 1/3. Another reason to exclude these morphologically complex structures is that they may not give good effective radius, velocity dispersion and density estimates. In Fig.\ref{total}(d), the remaining branch structures correspond well to the background integrated intensity.
For each structure, its velocity range and effective ellipse are used to extract the basic physical quantities based on the column density, temperature, optical depth cubes derived from the LTE analysis in Sec.\ref{cube}.

Branch structures are often contained within other branch structures. Some branch structures have similar central coordinates, scales, and morphology, thus they should be regarded as the same structure to avoid being repeatedly counted. Two branch structures with the area $s1$ and $s2$ ($s1$>$s2$) are considered repetitive if they meet the conditions: 1. The distance between their central coordinates is less than 1 beam size; 2. ($s1$-$s2$)/$s2$<1/3. The two clustering conditions can pack up the similar branch structures. Each clustering may contain multiple structures, and we keep only one of them in the subsequent analysis.
This step will exclude nearly half of branch structures. Thus the duplication of branch structures identified by the Dendrogram algorithm is a big issue,
it must be considered before making analysis for the identified structures.

\subsection{Column density}\label{column}
In this section, we derive the column density of the entire observed field using different methods to find the best estimates for the masses of the identified structures.

\subsubsection{Continuum emission}

\begin{figure*} 
\centering
\includegraphics[width=0.8\textwidth]{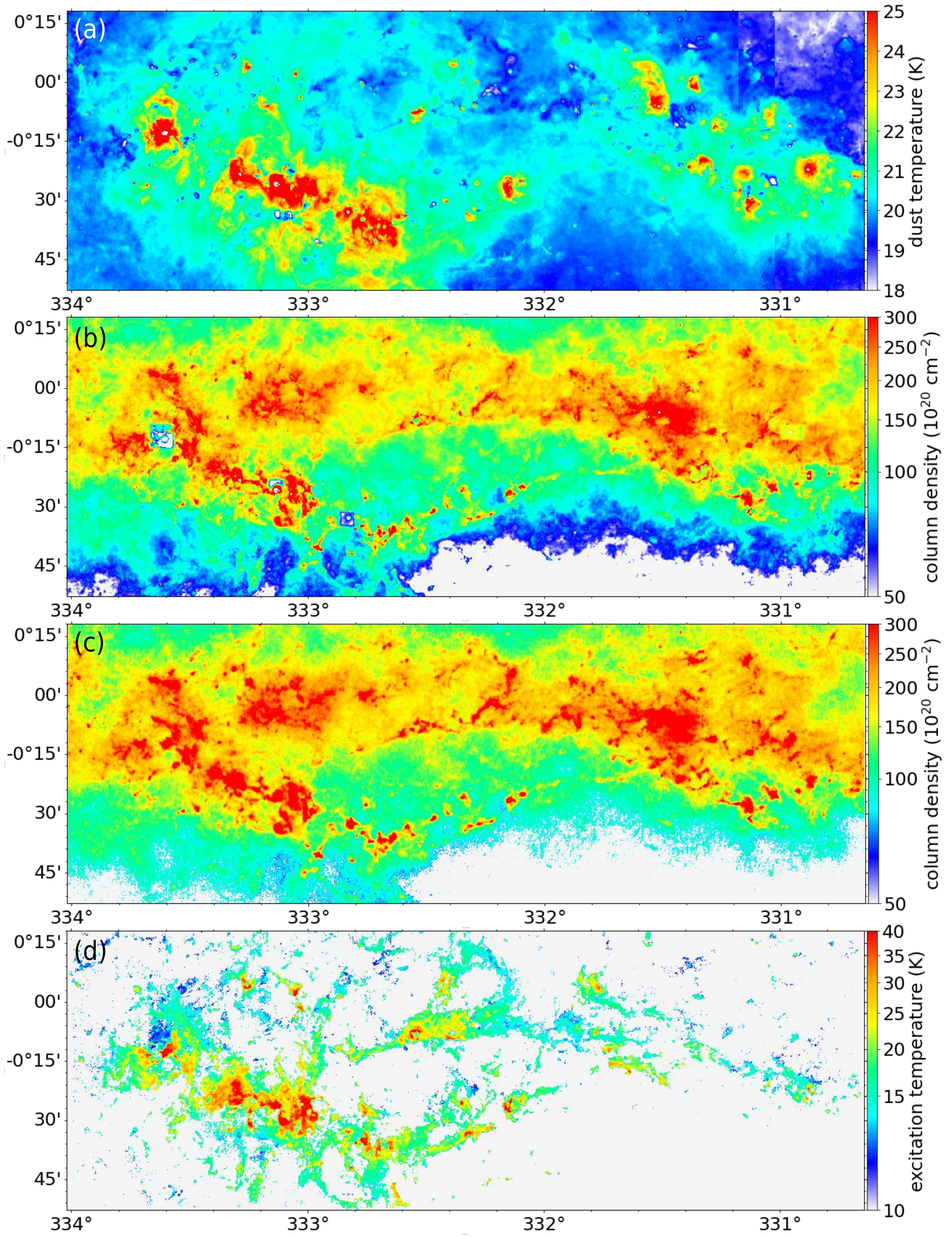}
\caption{
Temperature and column density maps of the entire field.
(a) and (b) Dust temperature and column density distributions in the G333 complex and the G331 GMC derived from Hi-GAL data processed by PPMAP; (c) Column density distribution in the G333 complex and the G331 GMC derived from ATLASGAL+Planck 870 $\mu$m data; (d) Excitation temperature distribution in the G333 complex derived from $^{12}$CO (3-2) emission by a LTE analysis in the velocity range [-60,-35] km s$^{-1}$.}
\label{continuum2}
\end{figure*}

Fig.\ref{continuum2}(a) and (b) present the dust temperature and column density maps derived from the Hi-GAL
data using the PPMAP procedure \citep{Marsh2015-454}. Since there are some missing values on the PPMAP column density map, we also produced the H$_2$ column density map using ATLASGAL+Planck 870 $\mu$m data following the formalism of \citet{Kauffmann2008-487}:
\begin{eqnarray}
N_{\rm H_2} & = &
  \displaystyle 2.02 \cdot 10^{20} ~ {\rm cm^{-2}}
  \left( {\rm e}^{1.439 (\lambda / {\rm mm})^{-1}
      (T / {\rm 10 ~ K})^{-1}} - 1 \right) 
   \left( \frac{\lambda}{\rm mm} \right)^{3} \nonumber \\
  & & \displaystyle
  \cdot \left( \frac{\kappa_{\nu}}{0.01 \rm ~ cm^2 ~ g^{-1}} \right)^{-1}
  \left( \frac{F_{\nu}}{\rm mJy ~ beam^{-1}} \right)
  \left( \frac{\theta_{\rm HPBW}}{\rm 10 ~ arcsec} \right)^{-2},
\end{eqnarray}
where $F_\nu$ is the flux density, 
$\theta_{\rm HPBW}$ is the beam FWHM, $\kappa_{\nu}=0.0185$ cm$^2$~g$^{-1}$ \citep{Csengeri2016-585}. 
Assuming a single dust temperature is a crude simplification, therefore we calculate N$_{\rm H_2}$ pixel-by-pixel by combining the ATLASGAL+Planck 870 $\mu$m flux map with Herschel dust temperatures derived with the PPMAP procedure. We only use pixels that are above the $\sim$5$\sigma$ noise level, $\sim$ 0.3 Jy/beam \citep{Urquhart2018-473}. From Fig.\ref{continuum2}(b) and (c), we can see
the column density derived from ATLASGAL+Planck 870 $\mu$m data and Herschel multi-wavelength data agree with each other, both in their spatial distribution and magnitude.

\subsubsection{Molecular line}\label{lte}
In this work, we focus on the G333 complex, and limit the velocity range of the $^{13}$CO emission to [-60,-35] km s$^{-1}$ \citep{Zhou2023-676}. To derive the column densities from the $^{13}$CO emission, we assume local thermodynamic equilibrium (LTE) and a beam filling factor of unity. Following the procedures described in \citet{Garden1991-374} and \citet{Mangum2015-127}, for a rotational transition from upper level $J+1$ to lower level $J$, we can derive the total column density by:
\begin{equation}
 N_{tot} =\frac{3k}{8\pi^3{\mu}^2B(J+1)}\frac{(T_{\rm ex}+hB/3k)\exp[hBJ(J+1)/kT_{\rm ex})]}{1-\exp(-h\nu/kT_{\rm ex})}
 \int{\tau{\rm dv}},
\label{density}
\end{equation}
\begin{equation}
\tau ={\rm-ln}[1-\frac{T_{\rm mb}}{J(T_{\rm ex})-J(T_{\rm bg})}],
\end{equation}

\begin{equation}
\int{\tau{\rm dv}} = \frac{1}{J(T_{ex}) - J(T_{\rm bg})} \frac{\tau}{1-e^{-\tau}}
\int{T_{\rm mb}{\rm dv}},
\label{tem}
\end{equation}

\begin{equation}
J(T) = \frac{h\nu/k}{e^{h\nu/kT}-1},
\label{j}
\end{equation}
where $B=\nu/[2(J+1)]$ is the rotational constant of the molecule, $\rm \mu$ is the permanent dipole moment ($\rm \mu= 0.112$ Debye for $^{13}$CO). $T_{\rm bg}=2.73$ is the background temperature, and $\int{T_{\rm mb}{\rm dv}}$ represents the integrated intensity. In the above formulas, the correction for high optical depth was applied \citep{Frerking1982-262,Goldsmith2008-680,Areal2019-36}.
Assuming optically-thick emission of $^{12}$CO emission, 
we can estimate the excitation temperature $T_{\rm ex}$ following the formula \citep{Garden1991-374,Pineda2008-679}
\begin{equation}
T_{\rm ex, 3-2} = \frac{16.6 {\rm K}}{{\rm ln}[1 + 16.6/ ( ^{12}T_{\rm peak, 3-2} + 0.038)]}, 
\label{tex}
\end{equation}
\begin{equation}
T_{\rm ex, 1-0} = \frac{5.53 {\rm K}}{{\rm ln}[1 + 5.53 / (^{12}T_{\rm peak, 1-0} + 0.818)]},
\label{tex}
\end{equation}
where $^{12}T_{\rm peak, 3-2}$ and $^{12}T_{\rm peak, 1-0}$ are the observed $^{12}$CO (3-2) and $^{12}$CO (1-0) peak brightness temperature. For the $^{13}$CO (2-1) transition, we do not have $^{12}$CO (2-1) data and we assume $T_{\rm ex, 2-1} = T_{\rm ex, 3-2}$. 

The distribution of the excitation temperature derived from $^{12}$CO (3$-$2) in Fig.\ref{continuum2}(d) is somewhat similar to the distribution of the dust temperature derived from Herschel data shown in Fig.\ref{continuum2}(a), especially in high-column density regions.
We transfer the column densities of $^{13}$CO to H$_{2}$ column densities by taking the abundance ratio X$_{\rm ^{13}CO}$ of H$_2$ compared with $^{13}$CO as $\sim 7.1 \times 10^5$ \citep{Frerking1982-262}.

\subsubsection{Column density cube}\label{cube}

A similar procedure as presented in Sec.\ref{lte} can be performed for each velocity channel in the $^{13}$CO (3-2) cube to obtain a column density cube, which allows to eliminate the effect of potential overlap of different velocity components on the mass estimation. 


\begin{figure*}
\centering
\includegraphics[width=0.9\textwidth]{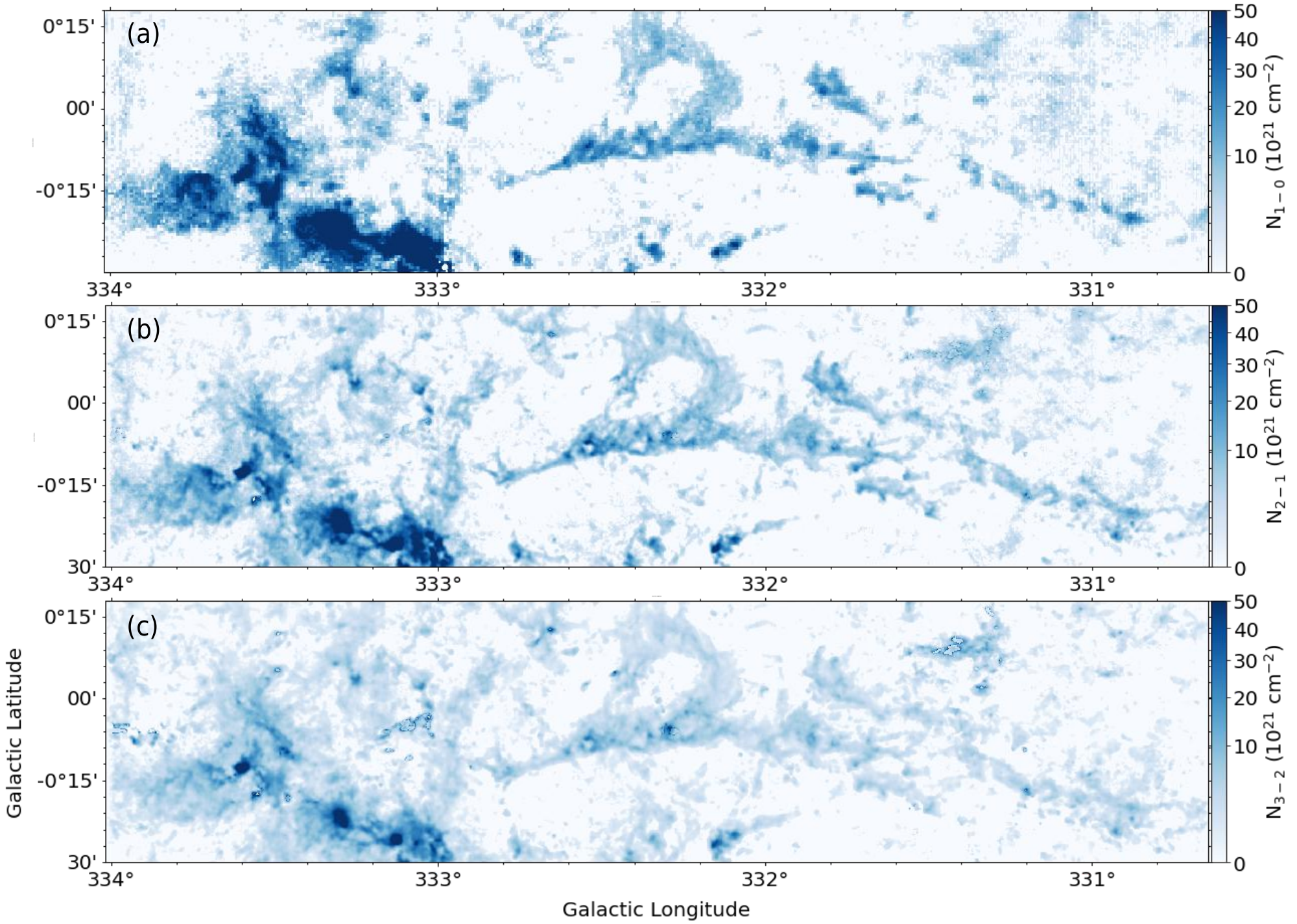}
\caption{H$_{2}$ column density distribution in the G333 complex derived from $^{13}$CO emission. (a) 1-0 transition; 
(b) 2-1 transition;
(c) 3-2 transition.}
\label{lte3}
\end{figure*}

\begin{figure}
\centering
\includegraphics[width=0.45\textwidth]{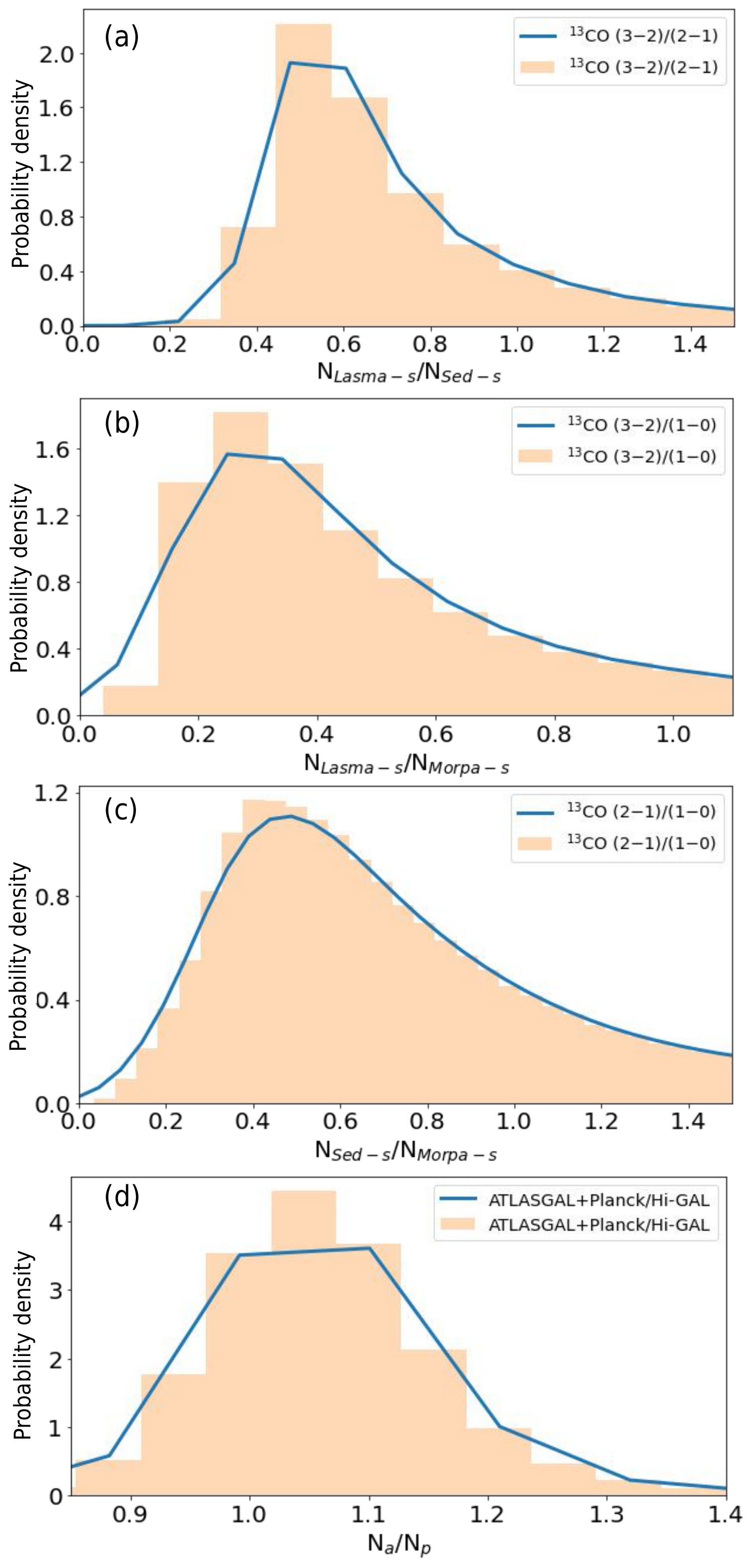}
\caption{Distribution of column density ratios: (a) derived from the $^{13}$CO (3-2) and (2-1) emission; (b) derived from $^{13}$CO (3-2) and (1-0); (c) derived from $^{13}$CO (2-1) and (1-0); (d) derived from ATLASGAL+Planck 870 $\mu$m and Hi-GAL data.
The probability density is estimated by the kernel density estimation (KDE) method.}
\label{p-ratio}
\end{figure}

\subsubsection{Column densities from different $^{13}$CO transitions} \label{co-ratio}
There are several factors that affect the mass estimate:
1.) the overlap of different velocity components; 2.) the observed molecular line transition; 3.) the choice between using molecular line  or continuum emission.
For the first factor, we have decomposed the velocity components in \citet{Zhou2023-676} and here we only focus on the peak3 component defined in \citet{Zhou2023-676} with the velocity range [-60,-35] km s$^{-1}$. 
For the second factor, 
\citet{Leroy2022-927} measured the low-$J$ $^{12}$CO line ratio R$_{21} \equiv$ $^{12}$CO (2-1)/$^{12}$CO (1-0), R$_{32} \equiv$ $^{12}$CO (3-2)/$^{12}$CO (2-1), R$_{31} \equiv$ $^{12}$CO (3-2)/$^{12}$CO (1-0), using whole-disk CO maps of nearby galaxies, and found galaxy-integrated mean values in $16\%{-}84\%$ of the emission of $R_{21} = 0.65\ (0.50{-}0.83)$, $R_{32}=0.50\ (0.23{-}0.59)$, and $R_{31}=0.31\ (0.20{-}0.42)$. Hence, the 3$-$2 transition of $^{12}$CO  resulted in significantly smaller column density estimates compared to the 1$-$0 transition. To check whether different transitions of $^{13}$CO  show a similar behavior in a Galactic giant molecular cloud, we collected $^{13}$CO (2$-$1) and $^{13}$CO (1$-$0) emission of the G333 complex, as described in Sec.\ref{archive}. 

In Sec.\ref{lte}, we have derived the column density of different transitions by a LTE analysis.
As shown in Fig.\,\ref{lte3}, the quality of the $^{13}$CO $J$=1$-$0 data is not as good as for $^{13}$CO $J$=2$-$1 and $J$=3$-$2, thus we set a column density threshold ($>$ 10$^{21}$ cm$^{-2}$) to exclude the unreliable low-column density emission from the $J$=1$-$0 transition before the comparison. 
Fig.\,\ref{p-ratio} shows the distribution of pixel-by-pixel column density ratios between different $^{13}$CO transitions. The peak values in the distributions are
\begin{equation} 
\frac{N_{\rm 2-1}}{N_{\rm 1-0}}\approx 0.5
\label{ratio}
\end{equation}
\begin{equation} 
\frac{N_{\rm 3-2}}{N_{\rm 1-0}}\approx 0.3
\label{ratio1}
\end{equation}
\begin{equation} 
\frac{N_{\rm 3-2}}{N_{\rm 2-1}}\approx 0.5.
\label{ratio2}
\end{equation}
Except for the slightly lower ratio between 2$-$1 and 1$-$0 transitions, the ratios calculated from different $^{13}$CO transitions are comparable with the results derived from $^{12}$CO emission in \citet{Leroy2022-927}, although the ratios from $^{12}$CO emission are derived from integrated intensity, rather than column density in the case of $^{13}$CO. 

\subsubsection{Non-LTE estimates}\label{non}
\begin{figure}
\centering
\includegraphics[width=0.5\textwidth]{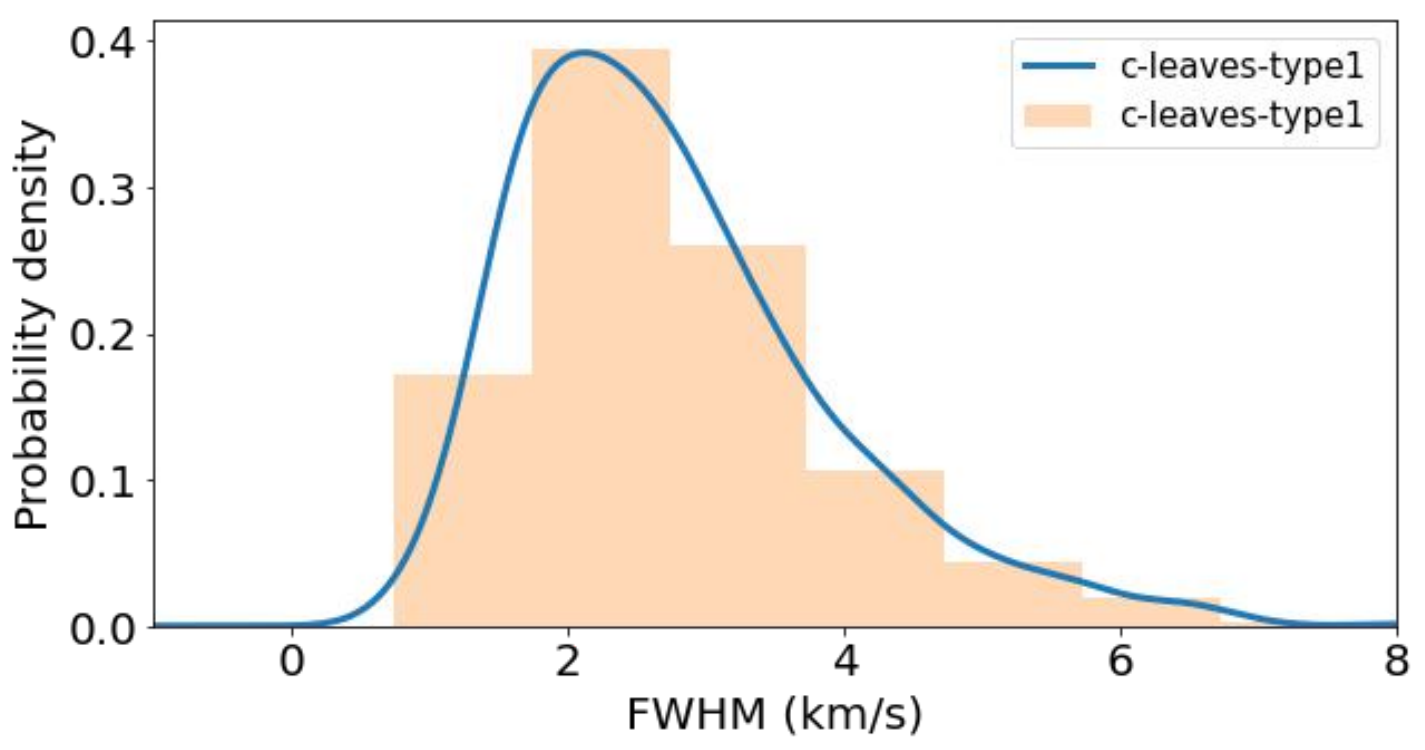}
\caption{Line-width distribution of all type1 c-leaves. The probability density is estimated by the kernel density estimation (KDE) method.}
\label{fwhm}
\end{figure}

\begin{figure*} 
\centering
\includegraphics[width=1\textwidth]{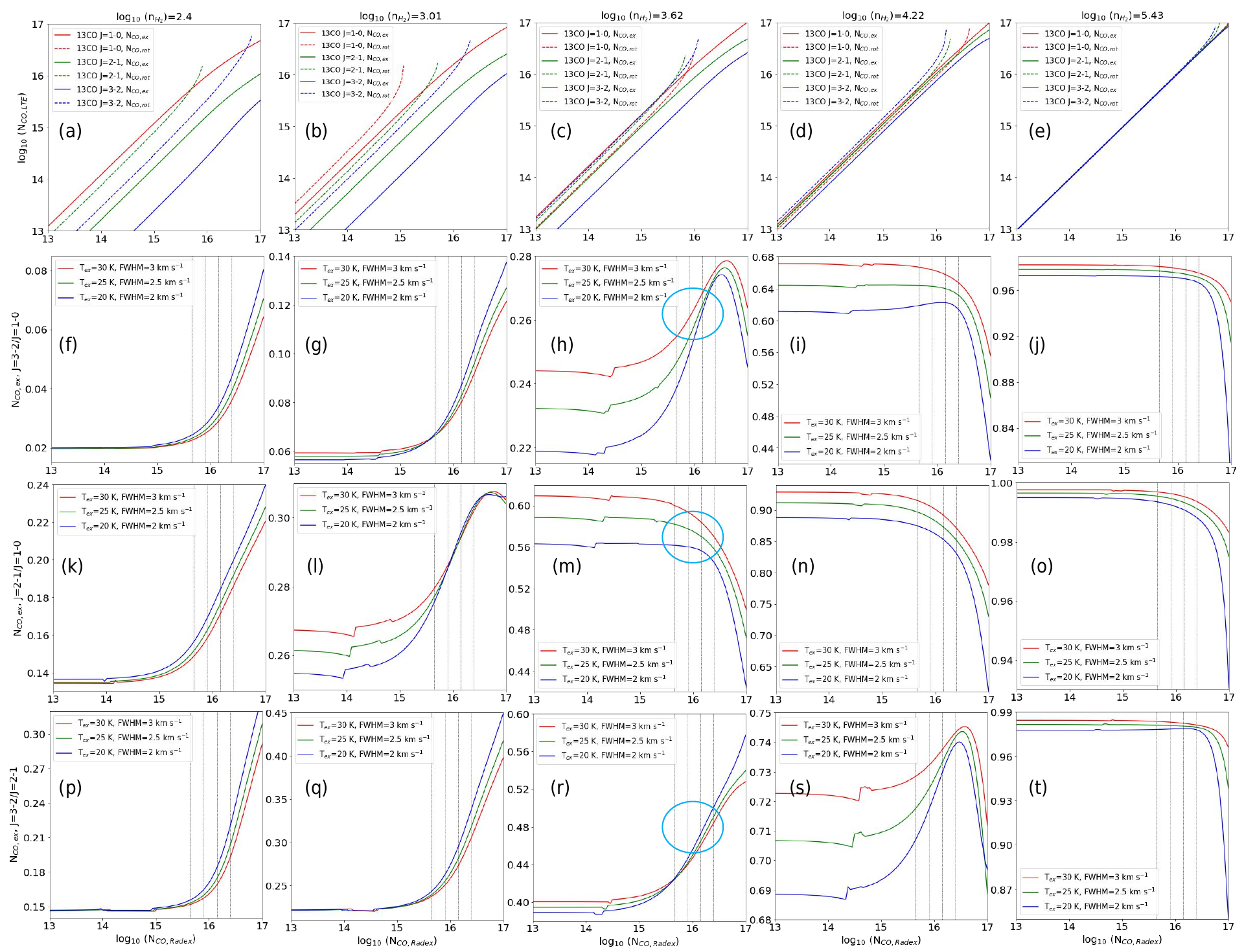}
\caption{The calculation results of RADEX. First row: correlation between RADEX input column densities with column densities of different $^{13}$CO transitions derived by LTE equations with T$_{\rm rot}$ and T$_{\rm ex}$ using T$_{\rm R}$ computed by RADEX for different volume densities; Second, third and fourth rows: Column density ratios of different $^{13}$CO transitions derived by LTE equations with T$_{\rm rot}$ and T$_{\rm ex}$ as a function of the RADEX column density input for different volume densities. Vertical lines mark the peak values of the $^{13}$CO column density derived from
$^{13}$CO J=3-2, J=2-1, J=1-0 and ATLASGAL+Planck 870 $\mu$m emission (from left to right) shown in Fig.\ref{continuum2}, Fig.\ref{lte3} and Fig.\ref{p-ratio}, here the abundance ratio X$_{\rm ^{13}CO}$ of H$_2$ compared with $^{13}$CO $\sim 7.1 \times 10^5$ is used. Cyan circles mark the ratios predicted by RADEX when the input $^{13}$CO column density takes the peak values of the column density derived by different methods in Sec.\ref{column}.}
\label{nlte}
\end{figure*}

The Non-LTE molecular radiative transfer algorithm RADEX was used to further test the above results: 1. The column density derived from $^{13}$CO $J=3-2$ transition is significantly lower than $J=1-0$ transition; 2. The ratios of the column density derived from different transitions of $^{13}$CO emission.
We use the following input parameters for RADEX: we take $T_{\rm kin}$=25 K as the kinematic temperature from Fig.\ref{continuum2}(a).
It is a mean value of the temperature in Fig.\ref{continuum2}(a) for the relatively high-column density regions in Fig.\ref{continuum2}(b), which covers the main emission of $^{13}$CO J=3-2.
As background temperature we use again $T_{\rm bg}$=2.73 K. A line-width of $\sim$2.5 km s$^{-1}$ is taken from the peak value of all type1 c-leaves, as shown in Fig.\ref{fwhm}. For the H$_{2}$ volume density log$_{10}$ (n$_{\rm H_{2}}$) and $^{13}$CO column density log$_{10}$ (N$_{\rm CO}$), we compute grids in the volume and column density range of
[2,6] and [13,17], respectively. Then we obtain the intensity $T_{\rm R}$ output from RADEX.
Assuming $T_{\rm ex}$= 25 K, using the equations listed in Sec.\ref{lte}, for the $J+1$ to $J$ transition,
the column density in the energy level $J$ can be calculated as
\begin{equation}
 N_{\rm J} = \frac{2J+1}{Q} N_{\rm tot, \rm CO} \exp[-\frac{hBJ(J+1)}{kT_{\rm ex}}],
\label{}
\end{equation}
where the partition function $Q$ is given by $kT_{\rm ex}/hB + 1/3$, $B$ is the rotational constant of the molecule. The rotation temperature $T_{\rm rot}$ can be estimated by the equation
\begin{equation}
\frac{N_{\rm j}}{N_{\rm i}} = \frac{g_{\rm j}}{g_{\rm i}}
\exp[-\frac{E_{\rm j}-E_{\rm i}}{kT_{\rm rot}}],
\label{}
\end{equation}
where N$_{\rm j}$ and N$_{\rm i}$ are the column densities of any two levels i and j of
statistical weights g$_{\rm j}$ and g$_{\rm i}$ and energies E$_{\rm j}$ and E$_{\rm i}$.
Using the equations listed in Sec.\ref{lte} again, now the column density N$_{\rm CO, rot}$ can be derived by T$_{\rm rot}$ and T$_{\rm R}$. We also derived the column density N$_{\rm CO, ex}$ by assuming T$_{\rm ex}$= T$_{\rm kin}$ = 25 K. Finally, N$_{\rm CO, rot}$ and N$_{\rm CO, ex}$ are compared with 
the $^{13}$CO column density N$_{\rm CO, radex}$ input in RADEX. As shown in Fig.\ref{nlte}, for N$_{\rm CO, ex}$, using the $^{13}$CO (3$-$2) emission together with the LTE equations indeed gives lower column density estimates than using the 2$-$1 and 1$-$0 emission, consistent with the results in Sec.\ref{co-ratio}. 
N$_{\rm CO, J=1-0, ex}$ provides upper limits of the column density derived by different $^{13}$CO transitions, thus it is used to calibrate the column density derived from $^{13}$CO (3-2) emission in this work.
Generally, for each transition, the column density N$_{\rm CO, rot}$ is higher than N$_{\rm CO, ex}$. The differences of N$_{\rm CO, rot}$ derived from different transitions are also smaller than that of N$_{\rm CO, ex}$. Moreover, N$_{\rm CO, rot}$ is closer to the fiducial N$_{\rm CO, J=1-0, ex}$ than N$_{\rm CO, ex}$. Therefore, using T$_{\rm rot}$ to derive the column density is better than using T$_{\rm ex}$.  
However,
both $J=2-1$ and $J=1-0$ data only cover part of the entire observed field, so that we cannot obtain the rotational temperature in the full region and do not use it here. 

In Fig.\ref{nlte}, we also investigate changes of the column density ratios N$_{\rm CO, ex}$ $J$=3$-$2/$J$=1$-$0,
$J$=2$-$1/$J$=1$-$0 and $J$=3$-$2/$J$=2$-$1 with the RADEX input $^{13}$CO column density N$_{\rm CO, radex}$ for different volume densities n$_{\rm H_{2}}$. Interestingly, when n$_{\rm H_{2}}$ is around $4.2 \times 10^3$ cm$^{-3}$ (the third column of Fig.\ref{nlte}), the predicted ratios are close to the values in Sec.\ref{co-ratio} for all three ratios. With $4.2 \times 10^3$ cm$^{-3}$ in the range of the typically volume densities of H$_{2}$ traced by  $^{13}$CO  \citep{Shirley2015-127,Liu2016-222,Schuller2017-601, Finn2021-917}, the column density ratios predicted by RADEX are consistent with the ratios derived using LTE from observations of  different $^{13}$CO transitions.

In summary, 
we divide the column density derived from $^{13}$CO J=3-2 emission by a correction factor 0.3 before converting to H$_{2}$ column density.


\subsection{Mass estimation}\label{mass1}
\subsubsection{Mass}
The mass of each identified structure is calculated by
\begin{equation}
M = \mu_{\rm H_2} {\rm m_H} \sum N{\rm(H_2)}(R\rm_{pixel})^2,
\label{c-eq8}
\end{equation}
where $\mu_{\rm H_{2}} = 2.8$ is the molecular weight per hydrogen molecule, 
$m_{\rm H}$ is the hydrogen atom mass, $ R\rm_{pixel}$ is the size of a pixel. The sum is performed within the elliptical cylinder in the column density cube. As described in 
Sec.\ref{Dendrogram}, the elliptical cylinder has a bottom area $A$=$\pi \times 2a \times 2b \times d^2$
and a height range [v$_{c}$-FWHM,v$_{c}$+FWHM], $a$ and $b$ are the long and short axes of the ellipse, here $d=3.6~{\rm kpc}$ for the distance to the G333 complex \citep{Lockman1979-232,Bains2006-367}.
Then the average surface density of each structure is calculated by $\Sigma$= $M/A$, and the average column density as $N$=$\Sigma/(\mu_{\rm H_2} {\rm m_H})$.

\subsubsection{Molecular line vs. continuum emission mass estimates}
\begin{figure}
\centering
\includegraphics[width=0.45\textwidth]{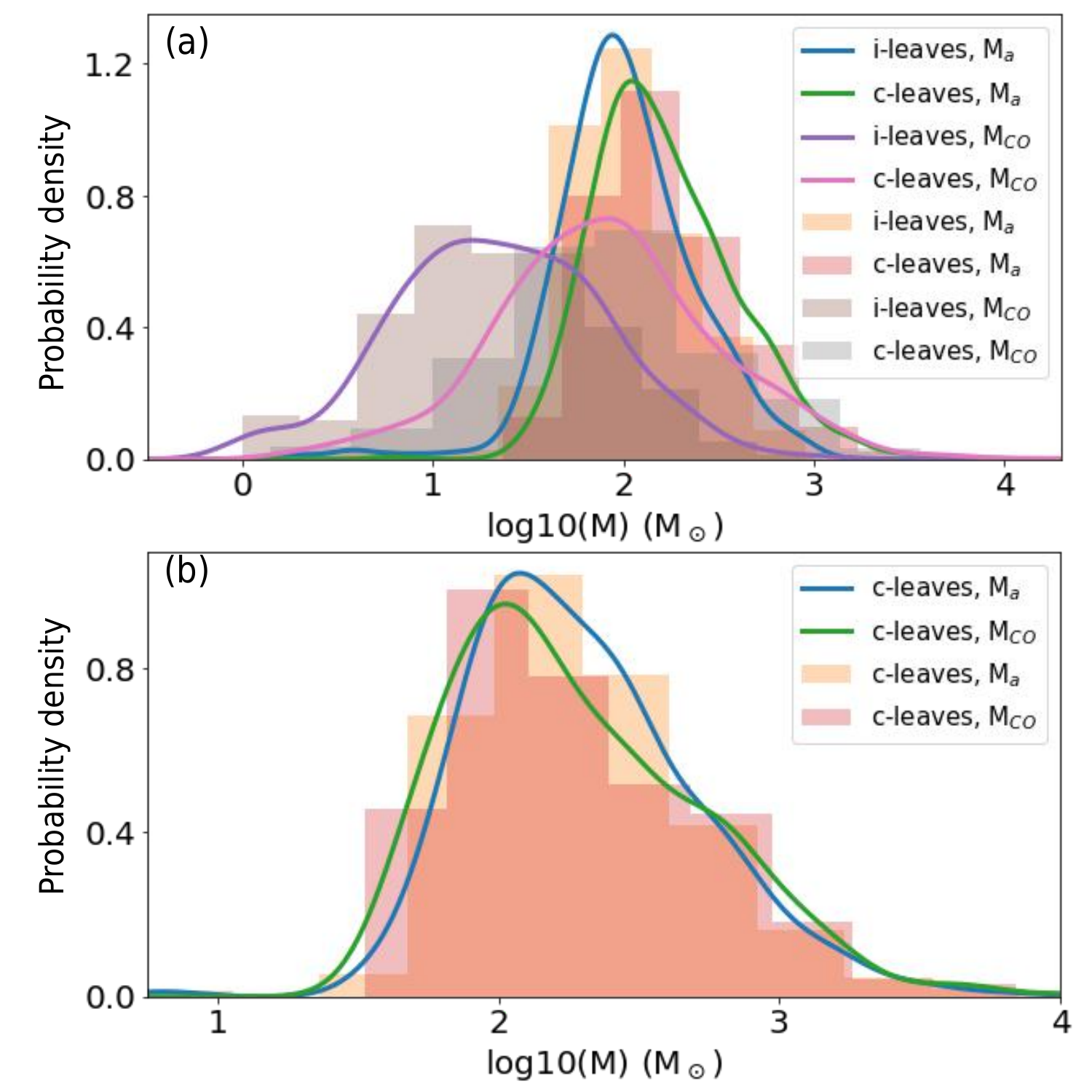}
\caption{M$_{\rm CO}$ and M$_{a}$ represent the mass derived from $^{13}$CO (3$-$2) emission and ATLASGAL+Planck 870 $\mu$m data, respectively.
(a) Mass distribution of all type1 leaf structures; (b) Mass distribution of type1 c-leaves satisfied the density condition >10$^{22}$ cm$^{-2}$.
The probability density is estimated by the kernel density estimation (KDE) method.}
\label{mass}
\end{figure}

As described in \citet{Zhou2023-676}, sub-regions 5 and 7 have significant overlap of different velocity components, thus they should be excluded for column density estimates based on continuum emission. In Fig.\ref{p-ratio}(d), the column density derived from ATLASGAL+Planck 870 $\mu$m data is comparable with that estimated from Hi-GAL data processed by the PPMAP procedure. 
As shown in Fig.\,\ref{total} and Fig.\,\ref{mass}(a), i-leaves are relatively low-column density structures distributed at the periphery, thus
we expect that they will be less massive on average than c-leaves, considering i-leaves and c-leaves structures have similar scales in Fig.\ref{basic}(a). However, in Fig.\ref{mass}(a), i-leaves and c-leaves show similar masses based on the continuum emission. 
In addition, the continuum mass distribution is relatively narrow, indicating that the contrast between high-column density and low-column density structures is not as clear as that derived from molecular line emission due to line-of-sight contamination. In Fig.\ref{mass}(b), we only consider the structures with mean column density greater than 10$^{22}$ cm$^{-2}$, 
now the distribution of the masses derived from molecular line emission is similar to that derived from continuum emission, after considering the mass correction factor 0.3. 
Therefore, in the subsequent analysis, we only adopt the masses estimated from molecular line emission.

\subsection{Virial analysis}\label{virial0}
Having measured the basic physical quantities of the identified structures, we can now investigate their physical properties.
\subsubsection{Virial parameter}\label{virial1}
\begin{figure}
\centering
\includegraphics[width=0.45\textwidth]{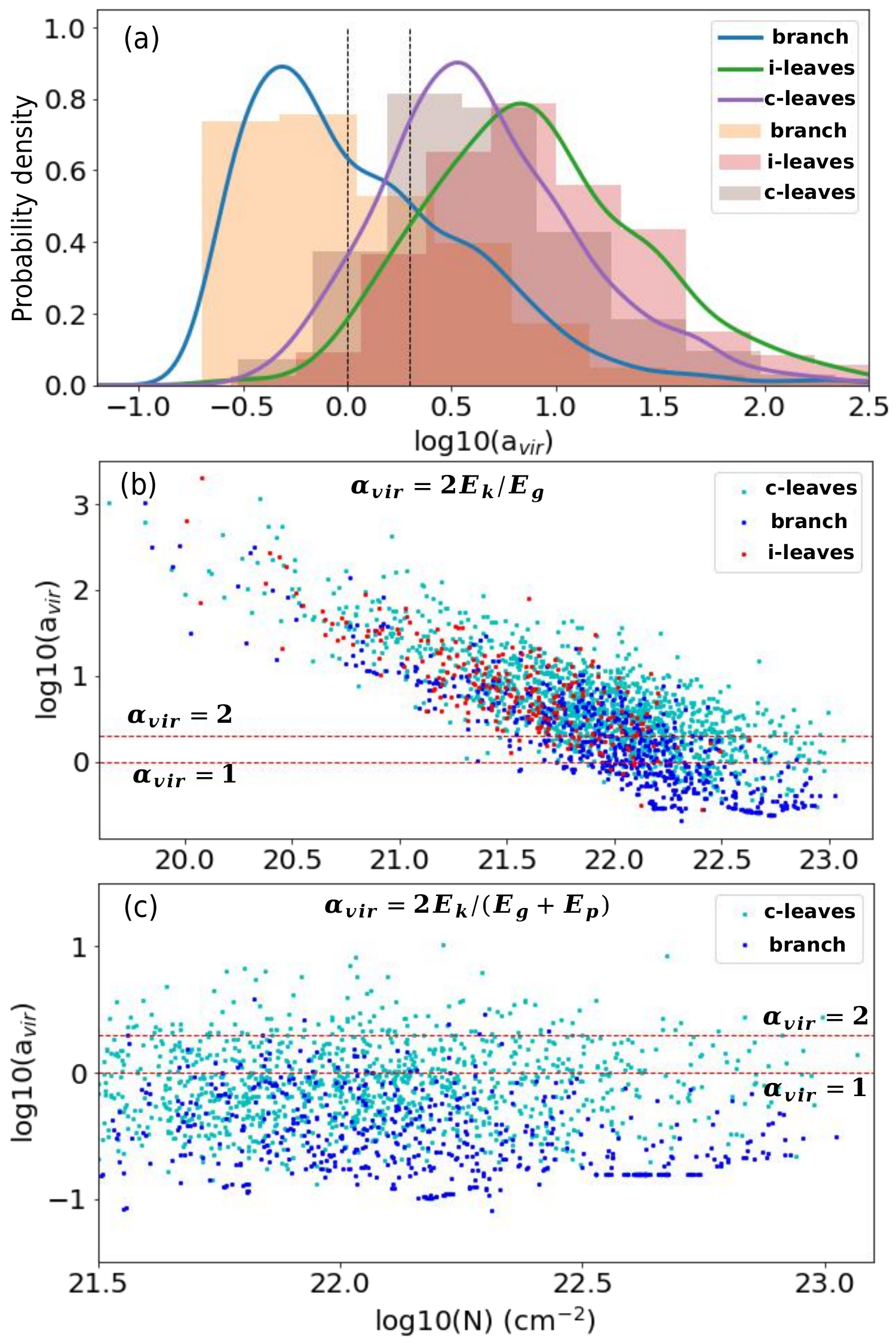}
\caption{Virial parameters of dense gas structures. (a) Virial parameters of all type1 and type2 structures. Dashed lines mark the positions $a_{vir}=1$ and $a_{vir}=2$. 
The probability density is estimated by the kernel density estimation (KDE) method; (b) Correlation between virial parameter $a_{vir} = 2E_{k}/E_{g}$ and average column density of structures; (c) Correlation between virial parameter $a_{vir} = 2E_{k}/(E_{g}+E_{p})$ and average column density above the threshold $\sim$3.2 $\times$ 10$^{21}$ cm$^{-2}$.}
\label{virial}
\end{figure}

To investigate the energy balance within the extracted structures, we determine the gravitational potential energy and internal kinetic energy to compute the virial parameter 
\citep{McKee1989-345,Bertoldi1992-395}: 
\begin{equation}
\label{eg}
E_{g} = - \frac{3}{5}a_{1}a_{2}\frac{ G M^{2}}{R}
\end{equation}
\begin{equation}
\label{ek}
E_{k} = \frac{3}{2} M \sigma_{\rm tot}^{2}.
\end{equation}
The factor $a_{1}$ measures the effects of a nonuniform density distribution and the factor $a_{2}$ the effect of the clump’s ellipticity. The virial parameter of each decomposed structure is calculated by:
\begin{equation}
\label{equ:virial}
\alpha_{\rm vir} = 2E_{k}/\left| E_{g} \right|
=\frac{5}{a_{1}a_{2}} \frac{\sigma_{\rm tot}^{2} R}{GM} ,
\end{equation}
with $\sigma_{\rm tot}$ = $\sqrt{\sigma_{\rm nt}^{2} + c_{s}^{2}}$ as the total velocity dispersion, $R$ the effective radius, $G$ the gravitational constant, parameter $a_{1}$ equals to $(1-k/3)/(1-2k/5)$ for a power-law density profile $\rho \propto r^{-k}$, and $a_{2} = (\arcsin e)/e $ as the geometry factor. Here, we assume a typical density profile of $k$ = 1.6 for all decomposed structures \citep{Butler2012-754,Palau2014-785,Li2019-886}. The eccentricity $e$ is determined by the axis ratio of the dense structure, $e = \sqrt{1 -(b/a)^{2}}$, $a$ and $b$ are the long and short axes of the ellipse. Non-magnetized cores with $\alpha_{\rm vir} <$ 2, $\alpha_{\rm vir} \sim$ 1 and $\alpha_{\rm vir} <$ 1 are considered to be gravitationally bound, in hydrostatic equilibrium and 
gravitationally unstable, respectively 
\citep{Bertoldi1992-395,Kauffmann2013-779}. 
Those with $\alpha_{\rm vir} >$ 2  could be gravitationally unbound, and are therefore either pressure-confined, or in the process of dispersal. 

Fig.\ref{virial}(a) shows the distribution of virial parameters for all identified structures.
We can see more than half of the leaf structures are gravitationally unbound and only a small fraction are in gravitational collapse. However, in \citet{Zhou2023-676}, we argue that molecular clouds in the G333 complex are in a state of global gravitational collapse, since the ubiquitous density and velocity fluctuations towards hubs imply the widespread presence of local gravitational collapse. Our previous work provides a more comprehensive approach to study the gas kinematics in the clouds. The dense structures in the clouds are connected to the surrounding environment through filaments, the gravitational state of the structures can be reflected by the velocity gradients along the filaments, which indicate the converging motions toward gravitational centers (hubs). In Fig.\,\ref{scale1}, except for the low-column density structures, most of structures have obvious correlation between velocity dispersion and column density, which indicates a gravitational origin of velocity dispersion, as discussed later in Sec.\ref{origin}. Thus most of structures must be gravitationally bound, even in a state of gravitational collapse.

That more than half of the structures have virial parameters larger than 2 in Fig.\ref{virial}, seems likely from not considering other forces that can bind the structures.
In the interstellar medium, each of the structures is embedded in larger scale structures and one can therefore assume that they are confined by various external pressures 
\citep{Keto1986-304, Lada2008-672, Field2011-416, Leroy2015-801, Kirk2017-846, Chen2019-875,Li2020-896}. To reconcile the gravitational collapse evidence from \citet{Zhou2023-676} with the apparent small fraction of gravitational unstable gas structures based on the classical virial analysis, we discuss in the following the additional effect of external pressure from ambient cloud structures.

\subsubsection{Pressure-confined hydrostatic equilibrium}\label{virial2}

Previous studies have suggested that the external pressure provided by the larger scale molecular cloud gas might help to confine dense structures in molecular clouds \citep{Spitzer1978,McKee1989-345,Elmegreen1989-338,Ballesteros2006-372,Kirk2006-646,Lada2008-672,Pattle2015-450,Kirk2017-846,Li2020-896}. The external pressure energy can be calculated by
\begin{equation}
\label{ep}
E_{p} = -4\pi P_{\rm cl} R^{3},
\end{equation}
and then the new estimation of the virial parameter is
\begin{equation}
\label{virp}
\alpha_{\rm vir} = 2E_{k}/(\left| E_{g} \right|+\left| E_{p} \right|).
\end{equation}

External pressure can have various origins, such as the turbulent pressure from the HI halo of molecular clouds \citep{Elmegreen1989-338}, the recoil pressure from photodissociation regions (PDRs) \citep{Field2011-416}, the infall ram pressure, or other intercloud pressures \citep{Bertoldi1992-395, Lada2008-672,Belloche2011-527,Camacho2016-833}. Here we mainly consider the external pressure from the ambient cloud for each decomposed structure using
\begin{equation}
\label{evirp}
P_{\rm cl} = \pi G \bar{\Sigma}\Sigma_{r},
\end{equation}
where $P_{\rm cl}$ is the gas pressure, $\bar{\Sigma}$ is the mean surface density of the cloud, $\Sigma_{r}$ is the surface density at the location of each structure \citep{McKee1989-345,Kirk2017-846}.
We assumed that $\Sigma_{r}$ is equal to half of the observed column density at the footprint of each decomposed structure \citep{Kirk2017-846,Li2020-896}. 

The total mass of the G333 complex is calculated by Eq.~\ref{c-eq8} as $\sim$1.03 $\times$ 10$^{6}$ M$_{\odot}$, comparable with the mass of $\sim$ 1.7 $\times$ 10$^{6}$ M$_{\odot}$ calculated in \citet{Miville2017-834} using CO (1$-$0) emission. Using the sum of all non-empty pixels as the total area, the mean surface density of the G333 complex is $\sim$ 0.071 g cm$^{-2}$ (or $\sim$340 M$_{\odot}$ pc$^{-2}$), corresponding to a column density of $\sim$1.5 $\times$ 10$^{22}$ cm$^{-2}$. The G333 complex is located in the molecular ring of the Milky Way, where the mean surface density is $\sim$200 M$_{\odot}$ pc$^{-2}$ \citep{Heyer2015-53}. Given that the G333 complex is the most ATLASGAL clump rich giant molecular cloud complex in the southern Milky Way, it should have a density higher than the mean value. The mean surface density of the G333 GMC in \citet{Miville2017-834} and \citet{Nguyen2015-812} are $\sim$ 120\,\msun\,pc$^{-2}$ and $\sim$ 220\,\msun\,pc$^{-2}$, respectively (see Sec.4.5 of \citet{Zhou2023-676} for more details). Adopting a conservative estimate, we take the value of $\bar{\Sigma} \sim$ 200 M$_{\odot}$ pc$^{-2}$ as a lower limit, corresponding to a column density of $\sim$9 $\times 10^{21}$ cm$^{-2}$. 

There are also many leaf structures with lower-column density, usually distributed in the periphery of the clouds.
Eq.~\ref{evirp} may be not valid for them, because $\Sigma_{r}$ in the equation is integrated from the cloud surface to depth $r$ of each structure in the cloud \citep{Kirk2017-846}. 
We therefore need to set a density threshold for the structures to determine whether they are eligible to be bound by the external pressure from the ambient cloud. 
In Fig.\,\ref{scale1}, high-column density and low-column density structures show different behaviors,
the turning point corresponds to a column density value $\sim$3.2 $\times$ 10$^{21}$ cm$^{-2}$, which will be used as the threshold.

We treat gravitationally unbound branch structures as c-leaves. Here we ignore i-leaves structures, they are isolated structures at the cloud periphery and are unlikely to be confined by external pressure from the ambient cloud. For c-leaves and branches, the proportion of the structures above the density threshold ($\sim$3.2 $\times$ 10$^{21}$ cm$^{-2}$) is 82.5\%. The proportions of $a_{\rm vir} = 2E_{\rm k}/E_{\rm g}<2$ and $a_{\rm vir} = 2E_{\rm k}/E_{\rm g}\geq 2$ are 45.3\% and 54.7\%, respectively. After considering the external pressure ($a_{\rm vir} = 2E_{\rm k}/(E_{\rm g}+E_{\rm p})$), their proportions become 93\% and 7\%. When accounting for external pressure, the majority of the structures are gravitationally bound, and susceptible to gravitational collapse, as shown in Fig.\,\ref{virial}(c). 
The peripheral structures, however, are at low-column densities and less bound by external pressure, therefore are likely to be dispersed by feedback. 

Here we do not consider the HI halo of molecular clouds and also ignore the external pressure exerted by HII regions. The latter should also be important due to the strong feedback in the G333 complex. But the energy injected into the clouds from HII regions might also destroy the clouds, therefore we only consider the external pressure from the ambient cloud in binding the structures. 
The rough estimate in this section shows the important role of the external pressure in confining the observed gas structures. 

\subsection{Scaling relation}\label{scaling}

\begin{figure*}
\centering
\includegraphics[width=0.925\textwidth]{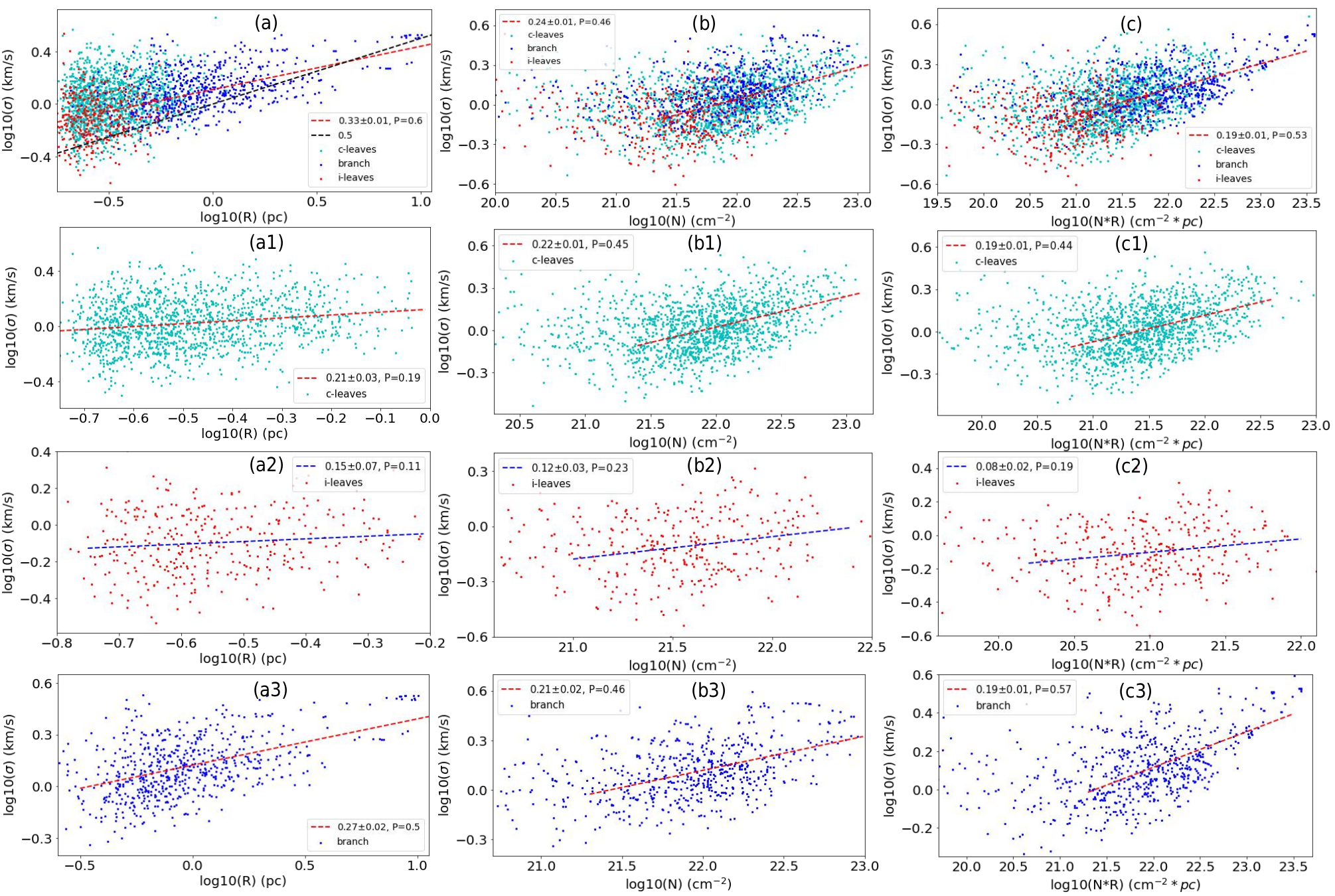}
\caption{Scaling relations of leaf and branch structures. (a) $\sigma-R$; (b) $\sigma-N$; (c) $\sigma-N*R$. $\sigma$, $R$ and $N$ are velocity dispersion, effective radius and column density of each structure, respectively. 'P' represents the Pearson coefficient.}
\label{scale1}
\end{figure*}

\begin{figure*}
\centering
\includegraphics[width=0.925\textwidth]{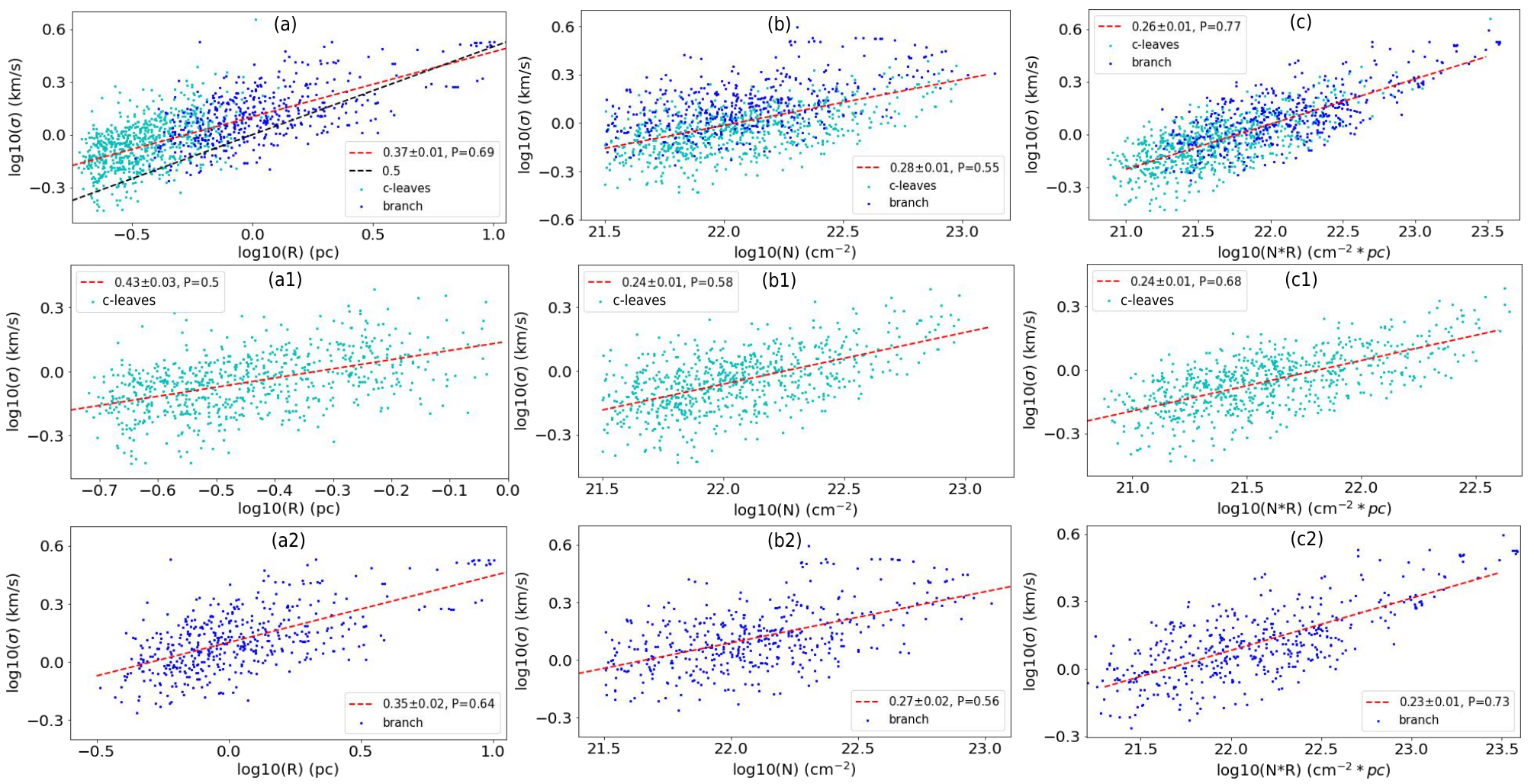}
\caption{Scaling relations of c-leaves and branch structures satisfied the conditions $N$>3.2 $\times$ 10$^{21}$ cm$^{-2}$ and $a_{vir} = 2E_{k}/(E_{g}+E_{p})<1$. (a) $\sigma-R$; (b) $\sigma-N$; (c) $\sigma-N*R$.}
\label{scale2}
\end{figure*}

\begin{figure}
\centering
\includegraphics[width=0.45\textwidth]{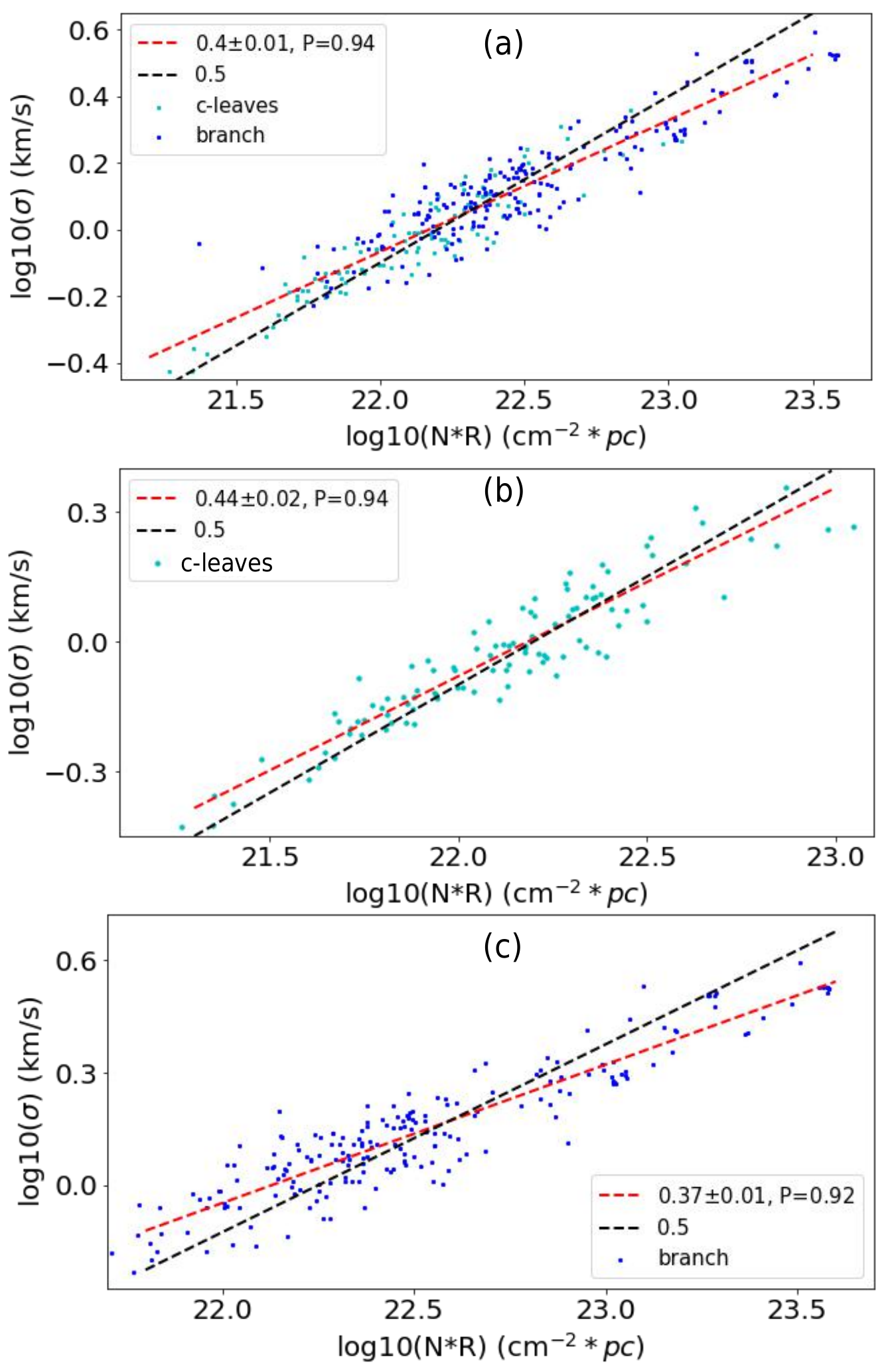}
\caption{Scaling relation $\sigma-N*R$ of c-leaves and branch structures satisfied the conditions $N$>3.2 $\times$ 10$^{21}$ cm$^{-2}$ and $a_{vir} = 2E_{k}/E_{g}<1$. 
}
\label{scale3}
\end{figure}

The physical states of the structures can also be reflected by the scaling relations.
Fig.\ref{scale1}(a) show the velocity dispersion-scale relations of i-leaves, c-leaves and branch structures. 
It appears that only branch structures show a clear correlation between velocity dispersion and scale. i-leaves roughly inherit the trend of branch structures' $\sigma-R$ relation extending from large to small scales, they can be barely linked behind branch structures in Fig.\ref{scale1}(a), although their velocity dispersion shows no significant correlation with scale, similar to c-leaves, which deviate more pronounced from the Larson-relation. 
The red dashed line, which is fitted to branch and i-leaves structures, has a gradient of 0.33$\pm$0.01.

Fig.\ref{scale1}(b) shows the velocity dispersion-column density relation. For c-leaves, there is a moderate correlation between velocity dispersion and column density, the Pearson coefficient is $\sim$0.45.
Interestingly, we can see different behaviors of high-column density and low-column density structures. For high-column density structures, the velocity dispersion and column density show a clear correlation, while low-column density structures do not. In recent simulations \citep{Ganguly2022arXiv,Weis2022arXiv}, dense structures roughly follow the Heyer-relation, and less dense structures show no trend with the column density, thus populate a low-density tail in the Heyer-relation, as shown in Fig.\ref{scale1}(c). For a more convenient comparison with $\sigma-R$ and $\sigma-N$ relations, we convert the Heyer-relation $\sigma/R^{0.5} \propto N^{0.5}$ to the form $\sigma \propto (R*N)^{0.5}$ \citep[Eq.\,3 in][]{Ballesteros2011-411}. Both of them should have a slope of 0.5.

From Fig.\ref{scale1}(a) and (b),
we conclude that the velocity dispersion of branch structures correlates with both scale and column density, that the velocity dispersion of c-leaves is only sensitive to column density, while the velocity dispersion of i-leaves has no significant dependence on either scale or column density.

The structures with column density $>$3.2 $\times$ 10$^{21}$ cm$^{-2}$ can be divided into two types: those that can collapse after adding external pressure ($a_{vir} = 2E_{k}/(E_{g}+E_{p})<1$, pressure-assisted collapse), and those that can collapse by self-gravity alone ($a_{vir} = 2E_{k}/E_{g}<1$, self-gravitating collapse). 
Now we have three structure sets: all identified structures, the structures in pressure-assisted collapse and the structures in self-gravitating collapse,
where the latter is a subset of the former.
The scaling relations of the three structure sets are shown in Fig.\,\ref{scale1}, Fig.\,\ref{scale2} and Fig.\,\ref{scale3}, respectively. For both leaf and branch structures, $\sigma-N*R$ has always a stronger correlation compared to $\sigma-N$ and $\sigma-R$. 
Moreover, the scaling relations show a stronger correlation and steeper slope when applied to self-gravitating structures, hence they best follow the Heyer-relation.



\subsection{Feedback}\label{feedback1}
\begin{figure*}
\centering
\includegraphics[width=0.925\textwidth]{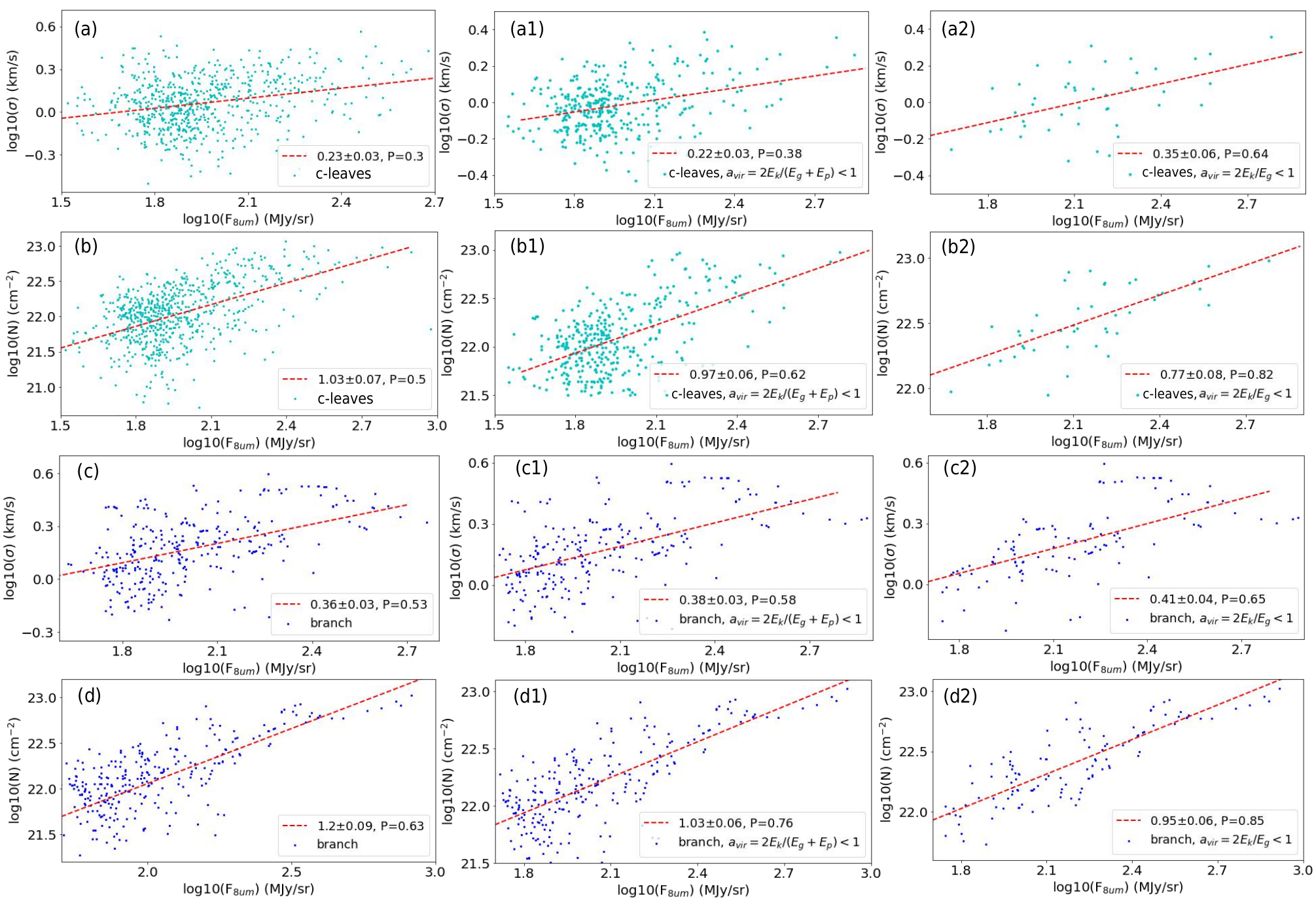}
\caption{Correlation between the 8 µm surface brightness and velocity dispersion and column density of the structures. 
First column: all type1 c-leaves and branch structures; Second column: type1 c-leaves and branch structures satisfied the condition $a_{vir} = 2E_{k}/(E_{g}+E_{p})<1$; Third column: type1 c-leaves and branch structures satisfied the condition $a_{vir} = 2E_{k}/E_{g}<1$.
}
\label{feedback}
\end{figure*}

In the study of the G305 molecular cloud complex, \citet{Mazumdar2021-650} argued that the 8~µm emission can be a good indicator of feedback strength. We calculated the average 8 µm surface brightness over each structure to measure the strength of feedback on each structure. In Fig.\,\ref{feedback}, the 8 µm surface brightness shows a strong positive correlation with column density for both c-leaves and branch structures, which might be an indication for triggering in the G333 complex. However, branch structures show a more obvious correlation between 8 µm surface brightness and velocity
dispersion than c-leaves, consistent with the results in \citet{Mazumdar2021-656}, implying that feedback has a greater impact on large-scale structures. 
The small-scale structures are embedded in large-scale structures, thus less affected by feedback. For large-scale structures, as shown in Fig.\ref{total}(c), the more evolved sub-region 1 and sub-region s2b are fragmenting into several pieces, potentially torn apart by the expanding HII regions. 
These results may explain why the velocity dispersion of branch structure has clear correlation with scale, leaf structure does not, as shown in Fig.\ref{scale1}. 
However, in Fig.\ref{scale2}, after filtering low-column density and gravationally unbound structures, 
velocity dispersion and scale of c-leaves appear to show better correlation than that in Fig.\ref{scale1}, the correlation between the 8 µm surface brightness and velocity dispersion of c-leaves is also improved in Fig.\ref{feedback}, which mean that leaf structure is also affected by feedback. 
Here we should remember that
there is a considerable overlap of the scales between leaf and branch structures, as described in Sec.\ref{Dendrogram}.

Our analysis is based on structural identification, when we say that feedback increases the density and velocity dispersion of the structures, provided that these structures can exist stably. $a_{vir} = 2E_{k}/(E_{g}+E_{p})<1$ and $a_{vir} = 2E_{k}/E_{g}<1$ structures can be more tenacious in feedback than other structures, and thus exhibit better scaling relations in Fig.\,\ref{scale2}, Fig.\,\ref{scale3} and Fig.\ref{feedback}. 
\citet{Dale2014-442} examined the effects, in simulations, of photoionization and momentum-driven winds from O-stars on molecular clouds, and found that feedback is highly destructive to clouds with lower mass and density, but have little effect on more massive and denser clouds.


\section{Discussion}

\subsection{The origin of velocity dispersion}\label{origin}

As discussed in \citet{Ballesteros2011-411,Traficante2018-473,Ballesteros2018-479,Li2023arXiv}, high-column density clumps or cores exhibit larger velocity dispersion than low-column density ones due to gas motions in gravitational collapse, as shown in 
Fig.\,\ref{scale1}(b) and Fig.\,\ref{scale2}(b), 
where the positive correlation between velocity dispersion and column density of c-leaves and branch structures 
indicates the gravitational origin of velocity dispersion. 
Combined with the discussions in Sec.\ref{scaling} and Sec.\ref{feedback1}, we conclude that both gravitational collapse and feedback contribute significantly to the velocity dispersion of large-scale structures. For small-scale structures, gravitational collapse is an important source of velocity dispersion, while the contribution of feedback needs more discussion in future work.

\subsection{The Heyer-relation in feedback}\label{heyer2}
In Sec.\ref{scaling}, self-gravitating structures can better fit the Heyer-relation. Considering that global collapse may lag behind the local collapse in the cloud \citep{Heitsch2008-674}, 
structures collapsing under self-gravity can be relatively independent of (or "decoupled from") the surrounding environment.
Then the explanations of the Heyer-relation in \citet{Heyer2009-699} and \citet{Ballesteros2011-411} can hold. Contrarily, for non-self-gravitating structures,
the exchange of energy with the surrounding environment will break the conversion between E$_{g}$ and E$_{k}$, thus breaking the Heyer-relation. 

\citet{Sun2018-860} measured cloud-scale molecular gas properties in 15 nearby galaxies,
and observed an excess in the velocity dispersion $\sigma$ at low surface density $\Sigma$ relative to the expected relation for self-gravity-dominated gas. This behavior leads to a shallower $\sigma-\Sigma$ relation in several galaxies, clearly deviating from a $\sigma-\Sigma^{0.5}$ relation extrapolated from the high surface density regime. One of their explanation for the deviation is that gas structures in the low surface density regime may be more susceptible to external pressure originating from the ambient medium or motions due to the galaxy potential, similar to our above explanation.

\subsection{Cloud disruption and collapse under feedback}
\citet{Peretto2023-525} performed an analysis of 27 infrared dark clouds (IRDCs) embedded within 24 molecular clouds. They found that the clumps are decoupling from their surrounding cloud and concluded that the observations are best explained by a universal global collapse of dense clumps embedded within stable molecular clouds, thus discovering direct evidence of a transition regime in the dynamical properties of the gas within individual molecular clouds.
As discussed in \citet{Heitsch2008-674,Ballesteros2011-411}, the decoupling may be due to the global collapse of the molecular cloud lagging behind the local collapse of dense clumps in the cloud. Invoking the notion of the "funnel" structure in PPV space \citep{Zhou2023-676}, a similar statement is that relatively small-scale hub–filament structures will have a more recognizable "funnel" morphology than large-scale ones due to their strong local gravitational field. 
Substructure s3a in the G333 complex is a vivid example of the decoupling highlighted in Fig.1 of \citet{Zhou2023-676}, which is collapsing into a hub-filament structure and separating from its surrounding environment.

One implication in the work of \citet{Peretto2023-525} is that star formation is likely to be mostly confined to parsec-scale collapsing clumps, also consistent with the results in \citet{Zhou2022-514,Zhou2023-676}. In our previous works,
for both ALMA and LAsMA data, most of the fitted velocity gradients concentrate on
scales $\sim 1$\,pc, a scale that is considered to be the characteristic scale of massive clumps \citep{Urquhart2018-473}. Velocity gradients measured around 1\,pc show that the most frequent velocity gradient is $\sim 1.6$\,km\,s$^{-1}$\,pc$^{-1}$. Assuming free-fall, we estimate the kinematic mass corresponding to 1\,pc is $\sim1190$\,M$_\odot$, which is also comparable with the typical mass of clumps in the ATLASGAL survey \citep{Urquhart2018-473}. Thus parsec-scale clumps are probably gravity-dominated collapsing objects.

The results in \citet{Zhou2022-514,Zhou2023-676,Peretto2023-525} show that the physical properties of parsec-scale clumps in two very different physical environments (infrared dark and infrared bright) are comparable. Thus feedback in infrared bright star-forming regions, such as the G333 complex, will not significantly change the physical properties of parsec-scale clumps, also consistent with the survey results that most Galactic parsec-scale massive clumps seem to be gravitationally bound no matter how evolved they are \citep{Liu2016-829, Urquhart2018-473, Evans2021-920}. 
Although the clumps are exposed to feedback and part of their velocity dispersion is due to feedback, as shown in Sec.\ref{feedback1},
the clumps are still self-gravitating sufficiently to continue their collapse, even after the lower density material has been disrupted and is being dispersed.
\citet{Watkins2019-628A} found that stellar feedback from O stars does not have much of an impact on the dynamical properties of the dense gas that has already been assembled, but does clearly modify the structure of the larger scale clouds.
The broken morphology of some very infrared bright structures in the G333 complex also indicates that the feedback is disrupting the molecular clouds.

The effects of feedback in star-forming regions can redistribute, disperse and enhance preexisting gas structures, and create new structures \citep{Elmegreen1977-214, Dale2007-375,Lee2007-657,Nagakura2009-399,Krumholz2014-243K,Fukui2021-73}.
According to the physical picture described in \citet{Zhou2023-676}, the hub-filament structures at different scales may be organized into a hierarchical system, extending up to the largest scales probed, through the coupling of gravitational centers at different scales. Large-scale velocity gradients always involve many intensity peaks, and the larger scale inflow is driven by the larger scale structure, implying that the clustering of local small-scale gravitational structures can act as the gravitational center on larger scale.
Given that the hierarchical hub-filament structures or the coupling of local gravitational centers in molecular cloud, and feedback does not impact much the dynamical properties of the dense gas, thus 
although the feedback disrupting the molecular clouds will break up the original cloud complex, the substructures of the original complex can be reorganized into new gravitationally governed configurations 
around new gravitational centers. This process is accompanied by structural destruction and generation, 
and changes in gravitational centers, but gravitational collapse is always ongoing.



\section{Summary}
We investigated the kinematics and dynamics of gas structures under feedback in the G333 complex. The main conclusions are as follows:\\

1. The dense gas structures were identified by the Dendrogram algorithm based on the integrated intensity map of $^{13}$CO (3$-$2). We obtained 3608 structures, their averaged spectra were extracted and fitted one by one. According to the line profiles, all averaged spectra were divided into three categories. Physical quantities of each structure were calculated based on their line profiles.

2. 
The column density of the entire observed field was derived from
ATLASGAL+Planck 870 $\mu$m data, Hi-GAL data, and different transitions of $^{13}$CO ($J$=1--0, 2--1 and 3--2). We investigated the column density ratios between them pixel-by-pixel, and found that the column density derived from ATLASGAL+Planck 870 $\mu$m data is comparable with that estimated from Hi-GAL data. Molecular line emission gives significantly lower column density estimates than those derived from the continuum emission. 
The peak values of the column density ratios between different transitions of $^{13}$CO emission are $N_{\rm 2-1}/N_{\rm 1-0} \approx$ 0.5, $N_{\rm 3-2}/N_{\rm 1-0} \approx$ 0.3, $N_{\rm 3-2}/N_{\rm 2-1} \approx$ 0.5. These ratios can be roughly reproduced by the Non-LTE molecular radiative transfer algorithm RADEX for typical volume densities of $\sim 4.2 \times 10^3$ cm$^{-3}$. Thus we adopted a correction factor of 0.3 to calibrate the column density derive from $^{13}$CO $J=3-2$ to be more representative of the total column density.

3. Classical virial analysis, suggesting many structures to be unbound, does not reflect the true physical state of the identified structures. After considering external pressure from the ambient cloud, almost all the structures with column density more than the threshold $\sim$3.2 $\times$ 10$^{21}$ cm$^{-2}$ are gravitationally bound, even undergoing gravitational collapse. 

4. The positive correlation between velocity dispersion and column density of c-leaves and branch structures reveals the gravitational origin of velocity dispersion. 

5. We use the average 8 µm surface brightness as indicator of feedback strength, which shows a strongly positive correlation with the column density of both c-leaves and branch structures. However, branch structures show a more significant correlation between 8 µm surface brightness and velocity dispersion than c-leaves, implying that feedback has a greater impact on large-scale structures. We concluded that both gravitational collapse and feedback contribute significantly to the velocity dispersion of large-scale structures. For small-scale structures, gravitational collapse is an important source of velocity dispersion, while the contribution of feedback needs more discussion in future work.

6. For both leaf and branch structures, $\sigma-N*R$ always has a stronger correlation compared to $\sigma-N$ and $\sigma-R$. 
The scaling relations are stronger, and have steeper slopes when considering only self-gravitating structures, which are the structures most closely associated with the Heyer relation.
However, due to the strong feedback in the G333 complex, 
only a small fraction of the structures are in a state of self-gravitational collapse. 

7. Although the feedback disrupting the molecular clouds will break up the original cloud complex, the substructures of the original complex can be reorganized into new gravitationally governed configurations 
around new gravitational centers. This process is accompanied by structural destruction and generation, 
and changes in gravitational centers, but gravitational collapse is always ongoing.

\begin{acknowledgements}
We would like to thank the referee for the detailed comments and suggestions that significantly improve and clarify this work.
This publication is based on data acquired with the Atacama Pathfinder Experiment (APEX) under programme ID M-0109.F-9514A-2022. APEX has been a collaboration between the Max-Planck-Institut für Radioastronomie, the European Southern Observatory, and the Onsala Space Observatory.
This research made use of astrodendro, a Python package to compute dendrograms of Astronomical data (http://www.dendrograms.org/). 
J. W. Zhou thanks  E. Vázquez-Semadeni for the helpful discussions. This work has been supported by the National Key R$\&$D Program of China (No. 2022YFA1603101). Tie Liu acknowledges the supports by National Natural Science Foundation of China (NSFC) through grants No.12073061 and No.12122307, the international partnership program of Chinese Academy of Sciences through grant No.114231KYSB20200009, and Shanghai Pujiang Program 20PJ1415500. 

\end{acknowledgements}

\bibliographystyle{aa} 
\bibliography{g333-1}

\end{document}